# Nanoscale roughness and morphology affect the IsoElectric Point of titania surfaces

F. Borghi[1], V. Vyas[1,2,†], A. Podestà[1]*, P. Milani[1]

*1) CIMaINa and Dipartimento di Fisica, Università degli Studi di Milano, via Celoria 16, 20133 Milano, Italy.*

*2) European School of Molecular Medicine (SEMM), IFOM-IEO Campus, Via Adamello 16, 20139 Milano, Italy.*

*† Present address: Institute of Material Sciences, University of Connecticut, 97 North Eagleville Road, Storrs CT, Connecticut, United States.*

* Corresponding author. E-mail: alessandro.podesta@mi.infn.it

## ABSTRACT

We report on the systematic investigation of the role of surface nanoscale roughness and morphology on the charging behaviour of nanostructured titania ($TiO_2$) surfaces in aqueous solutions. IsoElectric Points (IEPs) of surfaces have been characterized by direct measurement of the electrostatic double layer interactions between titania surfaces and the micrometer-sized spherical silica probe of an atomic force microscope in NaCl aqueous electrolyte. The use of a colloidal probe provides well-defined interaction geometry and allows effectively probing the overall effect of nanoscale morphology. By using supersonic cluster beam deposition to fabricate nanostructured titania films, we achieved a quantitative control over the surface morphological parameters. We performed a systematical exploration of the electrical double layer properties in different interaction regimes characterized by different ratios of characteristic nanometric lengths of the system: the surface rms roughness $R_q$, the correlation length $\xi$ and the Debye length $\lambda_D$. We observed a remarkable reduction by several pH units of IEP on rough nanostructured surfaces, with respect to flat crystalline rutile $TiO_2$. In order to explain the observed behavior of IEP, we consider the roughness-induced self-overlap of the electrical double layers as a potential source of deviation from the trend expected for flat surfaces.

# INTRODUCTION

Electrostatic interactions taking place at the interface of transition metal oxides (TMO) with water play a fundamental role in determining the behavior of systems and devices strategic for applications in biomedicine, catalysis, energy production/conversion, environmental remediation [1-3]. Biophysical phenomena such as the formation of bilayer membranes [4-6] or the adsorption and reorganization of proteins and cells at interfaces [7,8] depend upon the charging state of TMO surfaces in aqueous medium [8,9-12].

The charge of TMO surfaces in aqueous medium is mainly determined by two phenomena: protonation/de-protonation of surface hydroxyls [13-15], and adsorption of electrolyte ions onto the surface [16]. Two spatially defined regions of electric charge thus develop: a first compact layer of charge (Stern layer), closer to the solid surface and a few atomic sizes thick, including truly surface charges (originating in the amphoteric dissociation of surface groups) and surface-bound charges (adsorbed ions from the solution); a second diffuse layer of hydrated ions of both signs extends toward the bulk of the solution [17-19]. An electrostatic potential, solution of the Poisson-Boltzmann equation, exponentially decaying away from the surface, is associated to the overall charge distribution [19-21].

An important parameter to describe these electrostatic phenomena is the IsoElectric Point (IEP), which corresponds to the pH value at which the net charge of the compact layer is zero [22]. At IEP, also the $\zeta$ potential of the surface, which is responsible of the electrophoretic properties of particles in solutions [1,22,23], is zero, provided we identify the $\zeta$ potential with the potential at the boundary between the compact and the diffuse layers [22]. The Point of Zero Charge (PZC) corresponds to the pH required to have zero net surface charge. For an oxide surface without specific adsorption of ions (different from $H^+$ or $OH^-$) the IEP coincides with the PZC and, in particular, the $\zeta$ potential is negative for pH above the IEP, and positive below it [24,25].

When two interacting surfaces approach to a distance comparable or smaller than the typical screening length of the electrolytic solution (the Debye length, determined by the ionic strength of the solution), the overlap of the charged layers determines complex regulation phenomena [17] that are difficult to describe theoretically. In particular, when regulation phenomena occur, none of the following conditions, the constant surface charge or the constant surface potential, hold; these quantities become a function of the separation distance between the two interacting surfaces, or equivalently of the degree of overlap of the corresponding double layers. This brings the solution of the electrostatic problem far from the boundaries of the simplified linearized theory, which strictly holds only at low surface potential, large distances, and low ionic strength [19-21].

While significant insights have been obtained on the properties of the electric double layers formed between flat smooth surfaces [11,16,17,21], the case of rough surfaces still represents a severe challenge, hampering analytical, yet approximate, solutions of the double layer equations to be reliably obtained. Several authors have speculated that surface roughness may be responsible for discrepancies observed between experimental data and the predictions of the linearized DLVO theory; for example, a geometrical implication of surface corrugation is that the "average plane of charges", which produces the electrostatic double layer interaction, is shifted backwards with respect to the point of first contact between the surface and an incoming probe [26-31]. Despite the paramount importance of the explicit consideration of surface corrugation for the description of double layer electrostatic phenomena in real systems, and the significant theoretical efforts made to model electrostatic interactions at rough interfaces, the practical implementation of such models is still a land of pioneering studies, relying on approximated representations of rough morphology and/or on suitable approximation of the Poisson-Boltzmann equations. The interaction energy between mildly corrugated planes exhibiting periodic undulations (in the weak roughness regime, i.e. amplitude small compared to wavelength) has been calculated by means of Derjaguin approximation [32] by Tsao [33] and by Suresh et al. [34]. The surface element integration (SEI) technique allowed overcoming the limitations of the Derjaguin approximation when calculating the interaction energy between curves surfaces, modeled as a collection of convex and concave regions (spherical or sinusoidal bumps or depressions) with arbitrarily large curvatures (yet within the limits of the linearized PB equations) [35-39]. In these works an effort is made to relate the simplified topological model of surface roughness to statistical parameters that can be measured by an atomic force microscope (AFM), such as root-mean-square and other roughness parameters, specific area, etc.; moreover, it is recognized that the ratio of characteristic lengths of the system (Debye length, surface roughness, asperity separation…) influences the relative strength of different contributions to the interaction energy (van der Waals, electrostatic, Lewis acid-base acidity…). Duval et al. have explicitly included in their calculation of interfacial electrostatic interactions the charging mechanisms of the surfaces, developing a theoretical/numerical framework to account for local morphological (though calculations are implemented only for LEGO-like corrugated interfaces) as well as chemical heterogeneities of the surfaces. Their model takes into account the fine structure of the electrostatic double layer and boundary conditions beyond the limits of the linearized PB equations, allowing therefore to account for spatially-resolved charge regulation mechanisms and surface roughness effects [40]. Daikhin et al. have considered a statistical representation of surface morphology (in terms of height distributions) rather than on simplified geometrical constructions [41-43]; yet, their focus is limited to the calculation of some measurable electrochemical observables, typically the double layer capacitance. None of the works discussed so far present explicit calculations of the interaction force between rough surfaces in electrolyte solutions, and for this reason a direct application of theories for the analysis of experimental data acquired at complex rough interfaces is not straightforward.

Since most of the relevant biophysical phenomena cited above take place at the nanoscale, the characterization of charging mechanisms of nanostructured surfaces in electrolytic solutions and of the influence of the surface nanostructure is a necessary step towards the fundamental understanding and the effective exploitation of the role of nanostructured surfaces in tailoring and determining the functionality of the TMO interface with bio-objects [7-9].

A major problem hampering to reach a systematic and theoretically well-established description at the nanoscopic scale of interface charging is the lack of systematic experimental studies on double layer interactions at nanorough interfaces: in particular this is a consequence of the difficulty of preparing and characterizing, at the nanoscale, interfaces with controlled morphology, roughness, average slope, specific area, etc. Electrokinetic and electrophoretic measurements, potentiometric and calorimetric titration methods have been employed to characterize IEP and PZC of oxide particles in suspension [9,23,44-46], unfortunately these methods cannot provide quantitative local (i.e. at sub-micrometer scale) information of surface properties, and the application of these standard macroscopic techniques to surfaces in the form of thin films supported on solid substrates is problematic.

Here we report on the systematic and quantitative characterization of the role of nanoscale morphology on the charging behaviour of one of the most popular transition metal oxide surfaces: nanostructured titania. We have characterized IEP of nanostructured titania surfaces by direct measurement of the electrostatic double layer interaction in NaCl aqueous electrolyte using an atomic force microscope equipped with custom-made colloidal probes [47]. AFM is the technique of choice for sensing weak electrostatic forces (down to a few picoNewton) in solution, and has widely been employed to characterize double layer interactions (see, among many others references, Refs [25,29,48,49]); in those situations where surface roughness effects can be neglected, values of diffuse layer potentials measured by AFM and electrokinetic techniques have been found to be in good agreement [26,31,50,51].

Titania nanostructured films have been produced by supersonic cluster beam deposition (SCBD), a bottom-up approach providing a quantitative control over morphological nanoscale properties such as root-mean-square roughness, specific interfacial area, average surface slope [52-56]. Cluster-assembled titania surfaces has been recently demonstrated as a very reach playground to study the influence of nanostructure on proteins and cells [56-59].

In this manuscript we present experimental evidence of a marked dependence of the IEP of ns-$TiO_2$ surfaces on surface morphology, and we discuss our results on the basis of existing knowledge of the influence of surface morphology on double layer interactions; in the last part of the paper we consider the possibility that roughness-induced self-overlap of local diffuse layers acts as a potential source of deviation from the trend expected for flat surfaces.

# MATERIALS & METHODS

## Synthesis of nanostructured thin films by PMCS and reference substrates

A Supersonic Cluster Beam Deposition (SCBD) apparatus equipped with a Pulsed Micro-plasma Cluster Source (PMCS) has been used to deposit nanostructured titania (ns-$TiO_2$) films by assembling clusters produced in gas phase [52-54,60,61]. The PMCS operation principle is based on the ablation of a target rod by a helium or argon plasma jet, ignited by a pulsed electric discharge; the ablated species thermalize with helium or argon and condense to form clusters [60,61]. The mixture of clusters and inert gas is then extracted into the vacuum through a nozzle to form a seeded supersonic beam [54,62], which is collected on a set of round borosilicate glass coverslips (diameter 15 mm, thickness 0.13-0.17 mm) intercepting the beam in a deposition chamber. The clusters kinetic energy is low enough to avoid fragmentation and hence a nanostructured film is grown, leading to a highly porous, high-specific area material [55,56].

We deposited nine different ns-$TiO_2$ batches (samples SMP1-9 in Table 1, where the corresponding morphological parameters measured by AFM are also reported). In particular, ns-$TiO_2$ samples are characterized by thickness in the range 5-200 nm, rms roughness ($R_q$) ranging from 5 to 26 nm and specific area $A_{spec}$ from 1.2 to 1.8 (Table 1). Film roughness, specific area and the other chemico-physical parameters can be varied in a broad range by simply changing the thickness of the deposited films, without changing their surface chemistry [55]. Immediately prior to AFM characterization (morphological and electrostatic) ns-$TiO_2$ films have been thermally annealed for 2 hours at 250°C in ambient air, in order to remove organic contaminants and to recover the hydroxilated and hydrophilic surfaces.

The following substrates have been used as references to compare with the ns-$TiO_2$ film behavior: flat single-crystal <100> rutile $TiO_2$ (Sigma Aldrich), flat polycrystalline rutile $TiO_2$ and borosilicate glass coverslip (SLI Supplies). All the reference substrates were exposed to UV radiation for five minutes and then cleaned with ethanol and distilled water in order to remove contaminants from the surfaces. Borosilicate glass coverslips were used to realize a symmetrical system for DLVO measurements in order to characterize the net surface charge of the AFM probe at different pH (data presented in Supporting Information, section 5.1); to this purpose, in order to obtain surface properties comparable to those of the borosilicate glass colloidal probes, which undergo a thermal annealing above 750°C during production, borosilicate glass substrates were annealed at 600°C before characterization (it was not possible to anneal glass coverslips at higher temperature due to their tendency to bend significantly).

## Characterization of ns-$TiO_2$ films morphology

The surface morphology of ns-$TiO_2$ films was characterized in air using a Multimode AFM equipped with a Nanoscope IV controller (BRUKER). The AFM was operated in Tapping Mode, using rigid silicon cantilevers mounting single crystal silicon tips with nominal radius 5-10 nm and resonance frequency in the range 250-350 kHz. Several 2μm x 1μm images were acquired on each sample with scan rate of 1 Hz and 2048 x 512 points. The images were flattened by line-by-line subtraction of first and second order polynomials in order to remove artifacts due to sample tilt and scanner bow. From flattened AFM images root-mean-square surface roughness $R_q$ was calculated as the standard deviation of surface heights; specific area was calculated as the ratio of surface area to the projected area (more details on the calculation of morphological parameters are provided in Supporting Information, section 1). The film thickness was calculated by AFM, acquiring images across a sharp step produced masking the coverslip before the deposition.

## Characterization of electrostatic interactions by AFM

We have used a Bioscope Catalyst AFM (Bruker) to measure the electrostatic interactions between a colloidal probe and sample surfaces in electrolyte solutions with different ionic strength and pH. To this purpose force-distance curves (shortly force curves) were acquired by recording cantilever deflection versus piezoelectric translator displacement at the liquid/solid interface [49, 63,64]; ramp size was typically 1 μm (2048 points) with a scan rate of 1Hz. Samples were placed at the bottom of a petri dish filled by the electrolyte. The raw deflection signal from the detector in Volts was converted into a displacement in nm units multiplying by the deflection sensitivity factor (the inverse of the slope of the contact region of the force curve, acquired on a hard glass surface) [49], and then converted into force units in nN multiplying by the cantilever vertical force constant, calculated by thermal noise method [65]. The tip-sample distance D is calculated summing the cantilever deflection to the piezo displacement [63,64]. The long ramp size allows fitting and subtracting effectively an oscillating trend from force curves due to laser interference effects.

Force curves were acquired in aqueous solution (distilled Millipore water) with controlled ionic strength and pH, in the range 3-7 pH units at 20°C (see Supporting Information, section 3, for details). We have used a monovalent (1:1) electrolyte (NaCl) and a strong acid or base (HCl or NaOH) to change respectively the ionic strength and the pH of the solution [66,67]. NaCl electrolyte is an appropriate choice, because for low concentration ([NaCl] ≤ 0.1M) it is inert for $SiO_2$ [68] and $TiO_2$ [69-72] surfaces; it affects the value of the Ionic Strength but it does not change the value of the surface IEP. Setting the concentration of NaCl in pure water to 1mM (corresponding to $\lambda_D \approx 9.6$ nm) during experiments on ns-$TiO_2$ films allowed detecting weak electrostatic interactions with good signal-to-noise ratio for the reliable evaluation of surface charge parameters (this is critical in particular in the proximity of IEP, where net surface charge densities tends to zero); at the same time 1mM concentration is high enough to prevent modification of the ionic strength of the solution at the lowest pH values. For each sample 100 force curves were typically acquired in six different locations (separated by 100μm) in order to accurately characterize the Debye length and the charge densities of the surfaces (errors on Debye lengths and charge densities were calculated as described in the Supporting Information, section 3.2).

Colloidal probes provide a significantly enhanced signal-to-noise ratio compared to standard AFM tips and allow sensing the overall effects of nanoscale morphology, while a standard AFM tip with nanometer-sized apex would be sensitive to finer nanoscale fluctuations [29]. Moreover, colloidal probes determine a well-defined interaction geometry, allowing the use of simplified models to analyze data [21,29,48], where the radius of the probe can be set as a fixed and accurately calibrated parameter. We produced colloidal probes made of borosilicate glass following a novel protocol described in details in Ref. [47]. The probe size and its geometry are characterized by reverse AFM imaging of the probe on a MikroMasch TGT01 spiked grating (details are provided in Supporting Information, section 2).

Electrostatic and van der Waals forces in aqueous solution usually occur together and are considered additive in the Derjaguin-Landau-Verwey-Overbeek (DLVO) theory. In particular the interaction between a sphere and a flat surface is approximated by the following equations, valid for $D > \lambda_D$ [21,48,63,64,73,74]:

$$F_{DLVO}^{cc} = \frac{2\pi R\lambda_D}{\varepsilon\varepsilon_0}\left[2\sigma_S\sigma_T e^{-\frac{D}{\lambda_D}} + \left(\sigma_S^2+\sigma_T^2\right)e^{-\frac{2D}{\lambda_D}}\right] - \frac{AR}{6D^2} \qquad (1)$$

$$F_{DLVO}^{cp} = \frac{2\pi R\lambda_D}{\varepsilon\varepsilon_0}\left[2\psi_S\psi_T e^{-\frac{D}{\lambda_D}} - \left(\psi_S^2 + \psi_T^2\right)e^{-\frac{2D}{\lambda_D}}\right] - \frac{AR}{6D^2} \tag{2}$$

Here the superscripts *cc* and *cp* indicate constant-charge and constant-potential boundary conditions for the electrostatic contributions (first terms in Eqs. 1, 2, while the second terms represent the van der Waals force); the constant charge and constant potential conditions are typically well satisfied on insulating and conductive (metallic) surfaces, accordingly. R and $\sigma_T$ ($\psi_T$) are the radius and surface charge density (surface potential) of the sphere (the AFM probe), and $\sigma_S$ ($\psi_S$) is the surface charge density (surface potential) of the smooth (idealized) sample surface; ε is the dielectric constant of the medium (the aqueous electrolyte, we assume ε=78.54), $\varepsilon_0$ is the vacuum permittivity, $\lambda_D$ is the Debye length, i.e. the screening length of the electrolyte:

$$\lambda_D = \sqrt{\frac{\varepsilon\varepsilon_0 k_B T}{2e^2 I}} \tag{3}$$

where $k_B$ is the Boltzmann constant, T is the absolute temperature, e is the electric charge of the electron and I the ionic strength of the solution: $I = \frac{1}{2}\sum_i z_i^2 c_i$, $c_i$ and $z_i$ being the concentration (number of particles per unit volume) and valence of the i-th ionic species. The higher is the ionic strength, the more effective is the screening of electric fields in the solution. For 1:1 NaCl electrolyte with bulk concentration c=[NaCl], Eq. 3 simplifies to:

$$\lambda_D = 0.3/\sqrt{[NaCl]}\ \text{nm} \tag{4}$$

where the concentration of the salt is given in mol/l.

The Van der Waals force in Eqs. 1,2 depends on the Hamaker constant A of the surface/medium/probe system [73]. We have assumed for our experimental setup A=0.8 $10^{-20}$ J for borosilicate glass coverslip [29,49,75-77] and A=0.7 $10^{-20}$ J for ns-$TiO_2$ [78] (both against a borosilicate glass probe).

Potentials and surface charge densities in Eqs. 1,2 are related by the Grahame equation, which for a 1:1 electrolyte is [19]:

$$\sigma = \sqrt{8\varepsilon\varepsilon_0 k_B T c}\,\sinh\left(e\psi/2k_B T\right) \tag{5}$$

It should be noted that AFM tip senses the diffuse part of the electrostatic double layer [28,79], therefore surface charge densities $\sigma_S$ and $\sigma_T$ in Eqs. 1,2 must be identified with the surface charge density $\sigma_d$ of the diffuse layer, i.e. with the charge in the diffused layer projected on the outer Helmholtz plane; this charge density is equal in magnitude to the total charge density of the Stern layer : $\sigma_d = -(\sigma_0+\sigma_i)$, where $\sigma_0$ is the density of truly surface charges and $\sigma_i$ is the density of charges by ions from the electrolyte adsorbed (complexated) at the inner Helmholtz plane [17]. On amphifunctional surfaces, i.e. on surfaces where an electronic surface charging mechanism is present (as for example on bare, or partially oxidized, metallic surfaces), the previous equation must be changed in: $\sigma_d = -(\sigma_0+\sigma_i+\sigma_e)$, where $\sigma_e$ is the electronic surface charge density of the solid surface [11,16]. Our ns-$TiO_2$ however have a marked insulating character [80] and we will neglect in the following the $\sigma_e$ term. Under the assumption that the ions bind only to oppositely charged sites (energetically the most favourable option) it turns out that $\sigma_d$ represents a net surface charge density, being determined by the density of naked surface charges M-$O^-$ and M-$OH_2^+$ only, i.e. by those charges that are not neutralized by specifically absorbed electrolyte ions [22,46,79] (Supporting Information, section 4). At IEP $\sigma_d$ =0 while at PZC $\sigma_0$=0. AFM measurements can be used therefore to characterize IEP, not directly PZC, unless ion adsorption is negligible or symmetrical (indifferent electrolyte), in which case PZC=IEP.

The first terms of Eqs. 1,2 represent upper and lower limits for the general case of double layer interactions when charge regulation phenomena occur. We have tested the applicability of these simplified models to our systems, and concluded that the constant charge model is more appropriate to describe the experimental force data: the constant potential curves, built using potentials derived from charge densities according to Eq. 5 (in the limit of large distances, both *cc* and *cp* curves must overlap), systematically failed to reproduce the experimental data (details are provided in Supporting Information, section 3.1). Notice that while this suggests that the overlap of probe and sample double layers dos not lead to important regulation mechanisms, it does not imply that regulation phenomena are absent also *within* the double layer of corrugated ns-$TiO_2$ surfaces, as it is discussed later. For relatively large distances Eq. 1 simplifies to:

$$F_{DLVO}^{cc} = \frac{4\pi R\lambda_D}{\varepsilon\varepsilon_0}\sigma_S\sigma_T e^{-\frac{D}{\lambda_D}} - \frac{AR}{6D^2} \qquad (6)$$

Fitting average force curves with Eq. 6 provides the value of the charge densities product $\sigma_S\sigma_T$ and of the Debye length $\lambda_D$, the tip radius R being known from probe calibration (details in Supporting Information, section 2). In order to decouple from the fitted charge density product $\sigma_S\sigma_T$ the unknown contribution of the AFM borosilicate glass probe, we have characterized the net surface charge density of the borosilicate glass probe as a function of pH by recording force curve in aqueous electrolyte against a borosilicate glass smooth substrate, in order to realize a symmetrical system where $\sigma_S \approx \sigma_T$ and therefore $\sigma_T \approx \sqrt{\sigma_S \sigma_T}$ (Supporting Information, section 5.1, Fig. S8). This allowed in turn determining the absolute net surface charge density of flat crystalline $TiO_2$ and ns-$TiO_2$ surfaces.

Charge density products, rather than absolute charge densities, have been used to extrapolate $pH_{IEP}$ values, being this process based on the nullification of the prefactor of Eq. 6. To this purpose, all IEP values were extracted from $\sigma_S\sigma_T$ vs pH curves by interpolation between the closest experimental data with opposite sign, as shown in Supporting Information (section 5, data reported in Figs. S7B,S9right-S19right). In order to identify precisely the neighborhood of IEP on different surfaces, a few measurements at lower ionic strength ([NaCl]<$10^{-3}$mM) were typically performed, which reduces the electrostatic screening and increases the signal-to-noise ratio; these tests allowed identifying the pH values at which charge reversal takes place (Figs. S14right-S19right in Supporting Information, section 5.3). The determination of the $pH_{IEP}$ value is rather insensitive to the choice of the fitting model, being based on the nullification of surface charge product $\sigma_S\sigma_T$, rather than on the precise characterization of its magnitude in the neighborhood of the IEP. Overall, our setup is characterized by a sensitivity of about 2% in the determination of $pH_{IEP}$.

As part of the calibration of our experimental setup, in addition to determining the net surface charge density and IEP of the AFM probe, we have characterized the IEP of flat reference samples (Table 2; see Supporting Information, section 5.2, for details). Our experimental apparatus has proved to be accurately calibrated: the measured $pH_{IEP}$ values for borosilicate glass (silica-boron oxide mixture, annealed above 600°C), rutile single-crystal <100> and polycrystalline $TiO_2$ turned out to be in good agreement with the values reported in literature [9,72,83]. Robustness of the approach for the determination of $pH_{IEP}$ is witnessed also by the very good reproducibility of determination of IEP of the colloidal probe, despite the many different (chemically and morphologically) interfaces against which the probe has been used.

# RESULTS

## Surface Morphology of ns-$TiO_2$ Films

Fig. 1 shows representative AFM topographic maps of the ns-$TiO_2$ samples (both top- and 3-dimensional views), as well as single topographic profiles. The morphology of ns-$TiO_2$ films deposited by SCBD consists of a fine raster of nanometer-sized grains, with high specific-area, and porosity at the nano and sub-nanoscale depending on the film thickness [53-56], with grains diameter ranging from few nm up to 50 nm. Morphological parameters calculated from AFM topographies are reported in Table 1. The surface sections of Fig. 1 show nanometric pores of diverse depths and widths; an higher thickness means an increased geometrical accessibility of the pore, an increased local electric field strength around the sharpest asperities of the profile and a modification in the local surface charge distribution due to the overlapping, in the bottom and sides of the pore, of the diffuse double layers.

**Electrical double layer properties of rough ns-$TiO_2$ surfaces**

Fig. 2A shows average force curves for ns-$TiO_2$ films with roughness in the range 5-26 nm (SMP1-9, Table 1) at pH=5.4 and [NaCl]=1 mM (the ionic strength was kept constant through all the experiments, when not otherwise stated). At this pH all ns-$TiO_2$ surfaces are significantly charged. Fitting the curves shown in Fig. 2A by Eq. 6, we obtained the values of charge density and Debye length of all samples.

Fig. 2B shows the dependence on $R_q$ of the net surface charge density $\sigma_S$ of ns-$TiO_2$. The net surface charge density measured on the single-crystal rutile <100> $TiO_2$ surface, at the same pH, is also shown in Fig. 2B (empty square); this value represents a reference because the IEP of single-crystal <100> rutile is similar to those of rougher ns-$TiO_2$ surfaces (see below). In Fig. 2C we report the measured Debye lengths as a function of surface roughness of ns-$TiO_2$ films.

The trend of the charge density $\sigma_S$ of ns-$TiO_2$, which increases as $R_q$ increases up to a maximum value (for $R_q \approx 17$nm), then drops to values that are significantly lower than those of reference crystalline surface smaller values, is qualitatively and quantitatively counter-intuitive. Considering that the specific area of ns-$TiO_2$ samples increases (almost linearly - see Table 1) with $R_q$, we would expect on rougher surfaces a proportionally higher charge density with respect to the smooth rutile single-crystal <100> surface.

One would also expect that $\lambda_D$ does not depend on surface roughness, being a property of the bulk electrolyte, determined only by the ionic strength of the solution according to Eqs. 3,4. $\lambda_D$ is constant to a value $\lambda_D \approx 10$nm close to the one predicted by Eq. 4 for [NaCl]=1 mM only for $R_q < 20$nm, while on rougher samples $\lambda_D$ grows beyond 15 nm.

These experimental observations provide an indication that Eq. 6, which describes double layer interactions at smooth surfaces, may not provide an accurate description of charging and ionic re-distribution processes at rough surfaces. We have been therefore prompted by our data to consider the peculiar role of surface nano-morphology in electrostatic interactions between a microsphere and a rough surface, in the presence of an aqueous electrolyte.

Based on our observations and on previous reports [26-31] we have modified Eq. 6 in order to describe more accurately the probe-surface interaction force. Eq. 6 represents the approximated DLVO force in the case of a spherical colloidal micro-probe interacting with a smooth flat surface, such as for example the two crystalline reference rutile surfaces considered in this study. The situation when rough surfaces are involved, as in the case of ns-$TiO_2$ samples, is schematically represented in Fig. 3. A smooth object (the probe) is contacting the highest asperities of the surface of the nanostructured films; this is because the AFM probe is definitely too large to penetrate inside the surface nano-pores. The origin of distance axis in force curves corresponds to the point of first contact of the AFM tip with these protruding asperities, highlighted by the topmost red dash-dotted line in Fig. 3. Approximately, the separation between the actual contact line and the mid surface plane, represented by the lower dash-dotted line, is equal to $R_q$, the rms surface roughness. If we consider the mid-plane as an effective locus where all the electric surface charge is evenly distributed, it turns out that the distance axis for the double layer term in Eq. 6 must be shifted by $+R_q$ in order to recover an effective description of double layer interactions between a smooth and a rough surface. In other words, the average plane of charge in the case of corrugated surfaces is displaced backwards by $R_q$ (or by the sum of the $R_q$ of the two surfaces, in the case both are corrugated) with respect to the plane of first contact, located at the tops of surface asperities. We notice that while the shift of the distance axis does not change the value of IEP, determined by the zeroing of the product $\sigma_S\sigma_T$ in Eq. 6, it allows to evaluate more accurately the magnitude of such product. This is clear if we consider explicitly the effect of the shift of the distance axis on Eq. 6. If D is the apparent distance calculated from the point of first contact, the electrostatic force $F_{EL}$ at a distance $D+R_q$ from the mid plane is:

$$F_{EL} = \frac{4\pi R\lambda_D}{\varepsilon\varepsilon_0}\sigma_S\sigma_T e^{-\frac{D+R_q}{\lambda_D}} = \frac{4\pi R\lambda_D}{\varepsilon\varepsilon_0}\left(\sigma_S\sigma_T e^{-\frac{R_q}{\lambda_D}}\right)e^{-\frac{D}{\lambda_D}} \tag{7}$$

which can be written as a function of the apparent distance D as:

$$F_{EL}^{app} = \frac{4\pi R\lambda_D}{\varepsilon\varepsilon_0}\overline{\sigma_S\sigma_T}e^{-\frac{D}{\lambda_D}} \tag{8}$$

where

$$\overline{\sigma_S\sigma_T} = \sigma_S\sigma_T e^{-\frac{R_q}{\lambda_D}} \tag{9}$$

is an apparent charge density product ($\sigma_S$ reported in Fig. 2B is therefore an apparent charge density). Eqs. 7,8 show that when the distance axis is not shifted by $R_q$, the surface charge parameter extracted from the fit of Eq. 6 is exponentially underestimated by a factor depending on the ratio $R_q/\lambda_D$. Eq. 7 also predicts that the shift of the distance axis does not affect the Debye length.

The shift of the distance axis allows treating the rough surface as an effective smooth plane where the total surface charge is evenly distributed on the mid plane, which is approximately located a distance $R_q$ away from the surface peaks protruding towards the bulk of the electrolyte. A similar strategy has been adopted by the authors of Ref. [28], who pointed out that the potential at the outer Helmholtz plane of a rough gold surface (approximated by the ζ potential) can be rescaled by shifting the distance axis by an amount comparable to rms surface roughness; the authors applied to the electrostatic potential a correction similar to our Eq. 7. Similarly, Ducker et al. applied the same correction to extract the value of the surface potential of silica surfaces [29].

Fig. 4A shows the same force curves of Fig. 2A with corrected distance axes (all the distance axes of force curves shown from here on, and used to extract double layer parameters, have been shifted by $R_q$). Fig. 4B shows the corrected net surface charge densities $\sigma_S$ at pH 5.4 as a function of surface roughness. In Fig. 4B a clearer trend of the relative surface charge density vs $R_q$ is observed, with respect to Fig. 2B. $\sigma_S$ increases as $R_q$ increases: the increase is moderate for $R_q$<20nm; for $R_q$>20nm the increase is dramatic, and $\sigma_S$ of nanostructured samples is definitely much higher than that of smooth crystalline ones. The influence of surface roughness and specific area on charge density can be further appreciated in Fig. 5, showing the combined effect of pH and surface roughness ($R_q \geq$20nm) on the net surface charge density $\sigma_S$. As expected, $\sigma_S$ increases almost linearly as $|pH-pH_{IEP}|$ increases, due to the larger fraction of ionized surface groups. All samples (including SMP5, used for normalization) have similar IEP ($pH_{IEP}$~3.2, see later), i.e. at a given pH they should all be similarly charged. This is not the case, being evident that nanoscale morphology boosts the surface charge density in fact more than proportionally with respect to the increase in specific area.

Table 3 reports the value of IEP measured on different ns-$TiO_2$ surfaces. Fig. 6 shows the trend of IEP vs $R_q$ of ns-$TiO_2$ films. The observed shift of $pH_{IEP}$ is monotonic and seems to be only limited by the probed pH range: the loss of resolution in the measurement of $pH_{IEP}$ values on samples SMP5-8 is due to the fact that at these pH the AFM probe is almost neutral, therefore the force measured was very weak and the signal to noise ratio very low. The average force curves of each ns-$TiO_2$ sample at different pH, as well as the $\sigma_S\sigma_T$ vs pH curves, are reported in the Supporting Information (section 5.3, Figs. S11-S19). The difference between the $pH_{IEP}$s of ns-$TiO_2$ samples with lowest and highest surface roughness ($R_q$=5 nm and $R_q$=26 nm, accordingly) is remarkably more than two pH units and in particular the lower is the roughness of the ns-$TiO_2$ surface, the higher is the $pH_{IEP}$ value, with a monotonic trend towards the $pH_{IEP}$ of polycrystalline rutile $TiO_2$ ($pH_{IEP/polyTiO2}$= 6.28 ± 0.05) and anatase $TiO_2$ ($pH_{IEP}$ =6.1-6.3 [72]). This is consistent with the fact that the structure of ns-$TiO_2$ films is an amorphous matrix embedding rutile and anatase nano-crystallites [81,82], and that all the crystalline planes are likely randomly exposed. As $R_q$ increases, $pH_{IEP}$ monotonically decreases, reaching a value of 3.09 pH units. This value is close to that of flat single-crystal rutile <100>, which among different rutile crystallographic planes is the one exhibiting the lowest $pH_{IEP}$ [83].

# DISCUSSION

## Charging of metal oxide surfaces in aqueous electrolytes

The starting point in the discussion of experimental results is the consideration of the standard picture of surface charging of metal oxides in electrolytic solutions, which is generally attributed to the amphoteric character of surface hydroxyl groups [9-11,16,22,84]. Charging of the solid surface can be formally regarded as either a two-step protonation of surface M-$O^-$ groups, or equivalently as the interaction of surface hydroxyl M-OH with $OH^-$ and $H^+$ ions. In addition to association/dissociation of surface hydroxyls, also adsorption of anions $A^-$ and cations $C^+$ from solution to charged surface sites may take place. Details about the charging processes of oxide surfaces can be found in Supporting Information, section 4.

At the point of zero charge (PZC), the net electric charge at the solid/liquid interface is zero (the number of positively charged sites is equal to the number of negatively charged sites). This condition is achieved at a pH equal to [24,45]:

$$pH_{PZC} = 1/2\left(pK_1 + pK_2\right) - 1/2\log\left[\left(1 + K_+a_+\right)\left(1 + K_-a_-\right)\right] \qquad (10)$$

where $pK_i$=-$\log_{10}(K_i)$ (i=1,2,+,-), $K_{1/2}$, $K_{+/-}$ being the equilibrium constants for the association/dissociation reactions of the active species), and $a_{+/-}$ are the activity of cations and anions, accordingly (for 1:1 salt, like NaCl, $a_+=a_-\equiv a$).

At the Isoelectric Point (IEP), the net charge of the compact layer (i.e., also including the adsorption of anions and cations of the electrolyte) is zero. An expression for $pH_{IEP}$, similar to Eq. 10, has been obtained under the hypothesis that the slip plane coincides with the outer Helmholtz plane, i.e. the $\zeta$ potential is equal to $\psi_d$, the potential at the beginning of the diffuse layer [24]:

$$pH_{IEP} = 1/2\left(pK_1 + pK_2\right) - 1/2\left[\left(0.431\mathrm{e}^2\mathrm{N_s}/\mathrm{k_B C_1 RT}\right)a\left(K_- - K_+\right)\right]/\left[2 + \left(K_1/K_2\right)^{1/2} + a\left(K_- + K_+\right)\right] \quad (11)$$

In Eq. 11, $N_S$ is the total number of surface sites, $k_B$ is the Boltzmann constant, R is the universal gas constant, T is the absolute temperature, a is the bulk activity of NaCl, and $C_1$ is the capacity of the layer of ion pair localization, typically in the range 10-100 $\mu F/cm^2$. We have already stressed that IEP rather than PZC is characterized by AFM, because the AFM probe is sensitive to the overall charge of the compact Stern layer, or equivalently to the overall charge of the diffuse layer projected at the outer Helmholtz plane, which is equal and opposite, thanks to the electro-neutrality condition.

In order to get insights on how the evolving nanoscale surface morphology influences the IEP, we inquire the hidden role of morphological parameters in Eq. 10,11. We consider different possibilities, discussing them on the basis of our knowledge of charging mechanisms and of the physico-chemical properties of cluster-assembled titania.

Typically for smooth, flat surfaces in 1:1 aqueous electrolytes at low ionic strength, in the neighbourhood of the IEP/PZC (low surface potentials), one or more of the following conditions, leading to the equality $pH_{PZC}=pH_{IEP}$, are met:

i) Negligible ionic strength ($a\approx 0$);

ii) Negligible adsorption ($K_{+/-}\approx 0$);

iii) Symmetric adsorption ($K_+=K_-$).

According to Eqs 10,11, when conditions i)-iii) are met and $pH_{PZC}\approx pH_{IEP}$, changes of IEP can be due only to changes of pKs. When on the other hand conditions i)-iii) are not satisfied, also the activities $a_{+/-}$, as well as the equilibrium constants $K_{+/-}$, of electrolyte ions may couple to morphology and induce shift in the IEP. The picture is very complex because the failure of one or more of conditions i)-iii) can be itself determined by the evolving surface morphology. Equilibrium constants Ks depend on the atomistic properties of the surface, i.e. the density of active sites and the atomic neighbourhood of the active species (i.e. which atoms are bound to them, and by which kind of bond), and on the local electrostatic potential (i.e. on the local structure of the electrical double layer); ionic activities depend as well on the local electrostatic potential [46,85,86]. Clues to understand the morphology-driven variance of $pH_{IEP}$ and $pH_{PZC}$ of nanostructured oxide surfaces must be sought therefore in the morphology-induced modification of local surface chemistry and/or in the morphology-induced modification of the double layer structure. In the first case, the evolving morphology determines a change of IEP by directly modifying the local atomic environment of the active species (density of active sites, coordination, bonding); in the second case, the impact of the evolving morphology is more subtle and indirect, effectuating through the modification of the structure of the electrical double layer, i.e. through the modification of the electrostatic potential.

We will consider in the following firstly the possibility that morphology can change the local chemical environment of the active charge-determining surface species, and secondly the effect on electrical double layer. Before continuing, an important preliminary observation about the role of surface morphology must be done. IEP depends on the density of surface active sites rather than on their absolute number, i.e. IEP is an intensive surface property; this rules out the possibility that the observed shift of IEP on ns-titania towards more acidic pH is due to the increase of specific area on rough samples, i.e. to the capability of the surface to accommodate more (negative) charge due to the increased area, which would require more $H^+$ ions (lower pH) to achieve charge neutrality.

**Influence of nanoscale morphology on local chemical environment**

Several site-binding models [85,86] have been developed and proved to be effective in predicting the charging behaviour of oxide surfaces, and in particular the values of equilibrium constants and pKs, $pH_{PZC}$ and $pH_{IEP}$ values through Eqs. 10,11. According to these models, equilibrium constants depend on the atomic-scale environment and on the electronic properties of the surface sites (coordination, bond length, valence), as well as by the density of active sites, and on the electrostatic environment. Differences among IEP of different crystal faces of the same material can be readily accounted for by surface complexation models: individual surface planes of metal oxides, even in the absence of defects, typically possess several non-equivalent, differently coordinated oxygen atoms (singly, doubly, or triply coordinated), characterized by different activity coefficients.

A clear example of how surface structure affects PZC/IEP is the difference of $pH_{IEP}$ of different faces of rutile, recently determined by direct measurement of double layer forces by AFM [83]. A strong correlation of IEP with the density of cationic surface sites was demonstrated, the more acidic (with lowest $pH_{IEP}$ in the range 3.2-3.7) being the <100> surface of rutile. Polycrystalline surfaces of both rutile and anatase forms of $TiO_2$ possess the same PZC ($pH_{IEP}$≈6), resulting from the weighted average of the PZC of the single crystal faces.

Previous spectroscopic studies of electronic structure of ns-$TiO_2$ films produced using SCBD showed that $Ti^{3+}$ point defect states, related to oxygen vacancies and structural defects, are natively present in the material and relatively abundant; annealing at 250°C in presence of oxygen is effective in reducing the concentration of such defects [59,87]. Ns-$TiO_2$ films are mainly amorphous in nature, although both rutile and anatase nano-crystals are embedded in the amorphous matrix of the film [81,82]. There is evidence that the growth under sub-stoichiometric conditions in the cluster source favours the formation of rutile particles (typically for sizes below 5 nm). The differences in stoichiometry and crystalline phases of ns-$TiO_2$ films with respect to crystalline surfaces can account for static differences of PZC/IEP, but they could hardly account for the observed evolution of IEP with surface morphology. No evidence of any dependence of electronic and crystalline structure of ns-$TiO_2$ films on thickness and roughness has emerged from the mentioned previous spectroscopic studies.

Similarly to stoichiometry and crystalline phase, also the presence of chemical surface heterogeneities (including hydrophilic/hydrophobic nanoscale patches), partially penetrating the nanoporous matrix of the material, could in principle determine a change of IEP with respect to the pristine material; theoretical evidence has been recently provided of the direct influence of such surface chemical heterogeneities on electrostatic/electrokinetic interfacial properties [88,89]. However, the effects of such chemically different nanoscale domains on IEP should not evolve with rms roughness, but rather stay constant, as all sub-populations are equally amplified as the specific area increases.

A contribution from the $pK_{+/-}$ of the electrolyte ions could be expected from Eqs. 10,11, whenever the conditions i)-iii) are not satisfied. According to these equations preferential adsorption of anions leads to a decrease of IEP and increase of PZC (opposite trends are expected in the case of preferential adsorption of cations). On flat smooth interfaces, however, a slight predominance of one of the $K_{+/-}$ with respect to the other determines only small shifts of IEP/PZC by fractions of a pH unit, typically within the experimental errors, which are not comparable to the shift we have observed on nanostructured titania (more than 3 pH units, see Table 3). For example, in the case of $TiO_2$, $K_{Cl-}$ is reported to be slightly larger than $K_{Na+}$, but the maximum shift towards smaller values of $pH_{IEP}$ for variation of NaCl concentration over decades (from $10^{-3}$ M to $10^{-1}$ M) is only 0.8 pH units [24]. For this reason NaCl is generally considered as inert electrolyte towards smooth $TiO_2$ for low concentration ([NaCl] $\leq$ 0.1M) [69-72] (we verified this assumption by measuring double layer interactions on flat surfaces in the presence of ions at different concentrations, data not shown). We exclude therefore that small changes of the $pK_{+/-}$ for NaCl, due to different stoichiometry and crystalline phases of ns-$TiO_2$ with respect to crystalline $TiO_2$, can account for the observed marked shift of the IEP.

## Influence of nanoscale morphology on the structure of the electrical double layer

Ruled-out the direct influence of evolving surface morphology in changing the overall surface chemistry and therefore the pKs and the IEP of the system, we consider the possible effect of evolving morphology on the evolution of the structure of the electrical double layer, in particular on the electrostatic potential within the compact charge layer, which acts directly on pKs and activities; this could have potentially a very strong impact on the charging mechanisms of rough surfaces.

On rough surfaces, the double layer can be influenced by surface morphology, in particular by topological effects related to the local curvature, as well as to shadowing effects of surface charge and regulation mechanisms triggered by strong double layer overlap [17]. Standard DLVO theory developed for smooth surfaces and based on linearized Poisson-Boltzmann equations fails accounting for these topological effects. Although an approximate picture of the interfacial properties can be obtained by introducing the average plane of charge, i.e. by shifting the distance axis by $R_q$ towards larger distances, fine effects on double layer potential as well as counter-ion distribution related to surface morphology are not accounted for by this simple strategy. The anomalous behaviour of the Debye length shown in Fig. 2C can be an indication of this. Previous works have indeed suggested that surface morphology can affect the Debye length; on one hand, a surface-potential dependent Debye length, intended as an effective diffuse layer thickness, has been predicted for rough surfaces when non-linear Poisson-Boltzmann equations are considered [43]; on the other hand it has been recognized that on rough surface the electrostatic interaction has essentially three-dimensional components, therefore the extension of the electric field depends on surface morphology [40].

Recent works that have explicitly addressed the problem of solving the Poisson-Boltzmann equations in the case of rough (non-porous) surfaces [40-43] report that the properties of the double layer at a rough solid/liquid interface are mainly governed by the relative importance of ratios of the characteristic lengths of the electrode/electrolyte interface: $\lambda_D/\xi$ and $2R_q/\xi$, where $\xi$ is the lateral correlation length of the surface, i.e. the average peak to valley distance (see Supporting Information, section 1, for details) and $2R_q/\xi$ represents the average slope of the surface.

**Roughness-induced self-interaction of the electrical double layer**

Based on these works and on reports on charge regulation phenomena [1,17], we consider the idea of self-interaction of the double layer at nano-rough surfaces, i.e. the overlap of portions of the double layer pertaining to neighboring regions of the same surface; this effect is truly related to the corrugation of the surface, and in particular to the presence of contiguous regions with opposite slopes. A simplified case, that of two LEGO-like protrusions on a flat surface, has been previously addressed by numerical methods by Duval et al. [40]. Whenever double layers interact, either belonging to the AFM probe and the surface, or to adjacent surface regions, charge regulation phenomena occur, which, in the limit of strong overlap, may lead to severe distortions of the electrostatic potential and to the failure of the assumptions underlying the application of the linearized Poisson-Boltzmann equations.

By invoking a simplified geometrical model of the rough interface we suggest that the role of surface morphology is to enhance the self-overlap of double layer of neighboring surface regions. Figs. 7A,B show schematic representations of an average surface "pore". The pore is built by considering that, on average, peaks and valley across the surface are separated by a distance $\xi$ (the correlation length), and that about 70% of surface heights lies within a distance of $\pm R_q$ from the mid-plane, so that we may assume $2R_q$ as the average peak-to-valley separation. This picture is consistent with the fact that for gaussian surfaces the average surface slope is $2R_q/\xi$. In Fig. 7A the geometrical features of the average pore are highlighted. Assuming that the double layer stems perpendicularly from the surface up to a distance $\lambda_D$ from it (this cut-off is of course arbitrary, but does not influence the general conclusions of this reasoning), it turns out that because of the finite slope, double layers of adjacent walls overlap to some extent, the overlapping volume (an area in our 2-dimensional representation) being that of the quadrilateral enclosed by the dotted line in Fig. 7A. Qualitatively, the larger are $\lambda_D$ and surface slope, the stronger is the self-overlap of the double layer. The degree of morphology-induced self-overlapping of double layer on rough surfaces can be characterized by the fraction $\gamma$ of the double layer volume in each pore where overlap occurs. In our 2-dimensional representation $\gamma$ is the ratio of the area $\Sigma$ of the quadrilateral to the total area $\Sigma_0$ occupied by the double layer, i.e. $\gamma=\Sigma/\Sigma_0$. It turns out (details on calculations in the Supporting Information, section 6) that for $2R_q/\xi \leq 1$:

$$\gamma = \frac{(\lambda_D/\xi)(2R_q/\xi)}{2\sqrt{1+(2R_q/\xi)^2} - (\lambda_D/\xi)(2R_q/\xi)} \tag{12}$$

A similar formula holds for $2R_q/\xi>1$ (Eq. S9 in Supporting Information, section 6).

Eqs. 12 and S9 clearly show that the self-overlap of the double layer on nano-rough surfaces depends only on the ratios $\lambda_D/\xi$ and $2R_q/\xi$ of the characteristic electrostatic and morphological lengths (a similar scaling has been found by Daikhin et al. for the double layer capacitance [41-43]). In general, the degree of overlap inside each pore increases when the two ratios $\lambda_D/\xi$ and $2R_q/\xi$ increase. This can be also seen in Figs. 7A,B: upon increase of the slope at constant $R_q$, the overlap increases significantly. Eqs. 12,S9 also predict that for suitable combination of $\lambda_D$ and $\xi$ (relatively large $\lambda_D$ and small $\xi$) nearly complete overlap ($\gamma\approx1$) inside a pore can be reached. This condition is easily achieved on rough nanostructured surfaces, where pores of lateral half-width $\xi$ and vertical width $2R_q$ are decorated by smaller and smaller pores, whose local width and slope are typically higher than the mesoscopic quantities $\xi$ and $2R_q/\xi$. A schematic representation of the structure and sub-structure of the real pore of a nanostructured surface is shown in Fig. 7C (see also the topographic profiles shown in Fig. 1A,B,C), from which it is possible to infer that on rough nanostructured surfaces, the morphology-induced self-overlap of the electrical double layer can be dramatic. Overall, the roughness-induced self-overlap of the electric double layer brings the system far from the conditions when linearized PB equations hold, namely weak potentials and low ionic concentration, turning the interface into a strongly regulated one [17].

We think that regulation processes enhanced by double layer self-overlap can determine strong local gradient of surface potential and ionic concentration, leading to an increase of the net interfacial charge density $\sigma_d = (\sigma_0+\sigma_i)$ (what is measured by AFM). Redistribution of ions within the rough interface can be far from uniform, with a compression of the inner part of the diffuse layer inside the steepest and narrowest sub-pores, compensated by a depletion of the outer part, witnessed by an increase of $\lambda_D$ on rougher ns-$TiO_2$ samples (Fig. 2C).

Concerning the marked roughness-induced shift of IEP towards lower values, we can speculate mechanisms triggered by strong changes of the electrostatic potential due to double layer self-overlap and regulation effects. One such mechanism is the direct impact of the intense surface potential on the pKs, and therefore on the IEP [90], through Eq. 11; another mechanism is the rupture of the symmetry of cationic and anionic activities leading to a modification of the adsorption of electrolyte ions [16]. In the case of $TiO_2$, where a weak predominance of adsorption of anions with respect to cations has been reported [24], an enhancement of adsorption of $Cl^-$ anions would induce a downward shift of the IEP, according to Eq. 11.

The picture is further complicated by the fact that the investigated materials are porous in nature, being the result of random assembling of nanoparticles. The solid/liquid interface extends therefore inside the bulk material, inside nanopores, where extreme charge regulation effects may take place; the tail of the bulk double layer structure [91] can interfere with the outer double layer.

## CONCLUSIONS

This work represents a systematic effort aiming at mitigating the lack of experimental quantitative data on the effects of surface nanoscale morphology on the properties of electric double layers. The experimental approach we have adopted turned out to be very effective for the study of morphological effects on nanoscale interfacial electrostatic interaction. On one side, the use of SCBD technique for the synthesis of nanostructured titania films allowed to carry out a systematic investigation of the effects of nano-roughness on double layer properties thanks to the possibility of a fine control of morphological parameters; on the other side, operating an atomic force microscope in force-spectroscopy mode equipped with micrometer colloidal probes turned out to be effective in characterizing charging phenomena of nanostructured metal oxide thin film surfaces, a task which can hardly be accomplished by means of standard electrokinetic techniques, as well as by means of standard nanometer-sized AFM tips.

The most remarkable and novel result of our study is the observation of the shift of the IEP of cluster-assembled nanostructured titania by more than three pH units towards more acidic character with respect to reference crystalline surfaces, as the surface roughness increased from about 5 to 26 nm, values comparable to the Debye length of the electrolyte $\lambda_D$=9.6 nm. We have related the observed trend of IEP to the increasing importance of nanoscale morphology-induced self-overlap of the local diffuse layers, leading to strong charge regulation effects, local enhancement of surface potential and ionic concentration, and overall deviation from the trends expected for the linearized Poisson-Boltzmann theory. We propose a simple geometrical model for the self-overlap of the double layer, which highlights the importance of the ratios of characteristic lengths of the system (surface roughness $R_q$, correlation length $\xi$, and Debye length $\lambda_D$). Furthermore this model suggests that the competition of these lengths controls the properties of the double layer. In nanostructured interfaces all relevant morphological lengths are comparable to the electrostatic lengths $\lambda_D$ of the electrolytes; in particular, as $\lambda_D$ typically varies from a few angstroms to a few tens of nm, there will always be some surface structures of comparable size, in between the scale of single nanopores and that of mesoscopic structures of depth ~$R_q$ and width ~$\xi$.

The charging behavior of nanostructured surfaces may have important consequences for adsorption processes, as in the case of cell or protein-surface interactions. An incoming species, at a given distance from the surface (i.e. from the protruding asperities) of the order of one or two Debye lengths, will feel a reduced electric field compared to the case of interaction with a smooth surface, despite the fact that the surface is able to accommodate a greater amount of electric charge; this latter fact can be expected to play a role once the incoming species has approached to a distance comparable or smaller than the pore size, when the augmented local charge density and the dispersion forces will be felt directly and drive the final part of the adsorption process. The observed shift of the IEP on rough nanostructured titania films could potentially determine adsorption figures of proteins that markedly differ from those reported on smooth surfaces.

**Acknowledgements**
We thank C. Piazzoni and C. Lenardi for the deposition of ns-$TiO_2$ films.

**Supporting Information**

The following contents can be found in the Supporting Information document: 1) Characterization of surface morphology by Atomic Force Microscopy; 2) Characterization of colloidal probe radius; 3) Details on force curves and curve fitting procedures; 4) Charging of surfaces in liquid electrolytes; 5) Determination of charge density products and IEPs of reference systems; 6) Self-overlap of electrostatic double-layers: a simplified picture; 7) Bibliography.

**FIGURE CAPTIONS**

**Figure 1**. Top and 3-dimensional views of AFM topographic maps of ns-$TiO_2$ films. Thickness of ns-$TiO_2$ films is (A, D) 8 nm; (B, E) 50 nm; (C, F) 200 nm. Representative topographic profiles are superimposed to top-view maps.

**Figure 2.** (A) Average force curves at pH~5.4 and [NaCl]=1mM between the colloidal borosilicate glass probe and ns-$TiO_2$ films with different roughness. (B) The net surface charge density $\sigma_S$ of ns-$TiO_2$ versus pH, extracted from the best fit of average force curves by Eq. 6. For comparison, the net surface charge density of the reference <100> rutile $TiO_2$ surface is also shown. (C) Debye lengths $\lambda_D$ as a function of the surface roughness of ns-$TiO_2$ films extracted from the best fit of force curves by Eq. 6.

**Figure 3.** Schematic representation of the interaction geometry of a smooth micrometer-sized colloidal probe with a nano-rough surface. Red upper line: plan of first-contact, defined by the highest asperities; orange bottom line: mid-plane, or effective plane of charges. On average, the distance between the two planes is equal to $R_q$, the rms surface roughness.

**Figure 4.** (A) Average force curves at pH~5.4 and [NaCl]=1mM between the colloidal borosilicate glass probe and ns-$TiO_2$ films with different roughness with corrected distance axis (i.e. positively shifted by $R_q$, see main text for details). (B) The net surface charge density $\sigma_S$ of ns-$TiO_2$ versus pH, extracted from the best fit of force curves by Eq. 6 after correction of distance axis. For comparison, the net charge density of the reference <100> rutile TiO2 is also shown.

**Figure 5.** Evolution of the net surface charge density $\sigma_S$ with pH for ns-$TiO_2$ films with increasing roughness ($R_q \geq 20$nm; all films have similar IEP, see Table 3 and Figure 6).

**Figure 6**. $pH_{IEP}$ of ns-$TiO_2$ samples with different rms roughness; for comparison, $pH_{IEP}$ of flat single-crystal and polycrystalline rutile $TiO_2$ samples are shown.

**Figure 7**. Schematic representation of the self-overlap of electrical double layers taking place at corrugated interfaces. A simplified double layer extending to a distance $\lambda_D$ into the bulk of the electrolyte is shown. Surface pores are characterized by half-width $\xi$, height $2R_q$, and slope $2R_q/\xi$. A,B) Two pores with same height $2R_q$, same double layer depth $\lambda_D$, but markedly different slope. C) A "real" surface pore of a cluster-assembled nanostructured surface in aqueous electrolyte: pore structure is statistically scale-invariant, replicating at small scales.

**TABLES**

**Table 1.** Morphological parametrs of ns-$TiO_2$ samples measured by AFM.

| ***Ns-TiO$_2$ sample*** | ***Thickness (nm)*** | ***Roughness $R_q$ (nm)*** | ***Specific Area $A_{spec}$*** | ***Correlation length $\xi$ (nm)*** | ***Slope $2R_q/\xi$*** |
|---|---|---|---|---|---|
| SMP 1 | 7.7 ± 1.6 | 4.9 ± 0.1 | 1.19 ± 0.01 | 16.2 | 0.605 |
| SMP 2 | 31.4 ± 1.2 | 10.4 ± 0.7 | 1.21 ± 0.1 | 42.0 | 0.495 |
| SMP 3 | 33.9 ± 3.4 | 14.9 ± 0.2 | 1.41 ± 0.02 | 37.1 | 0.803 |
| SMP 4 | 50.5 ± 3.9 | 17.2 ± 0.1 | 1.56 ± 0.09 | 41.0 | 0.839 |
| SMP 5 | 62.0 ± 4.8 | 19.2 ± 0.4 | 1.61 ± 0.02 | 43.3 | 0.886 |
| SMP 6 | 96.5 ± 7.6 | 20.6 ± 0.1 | 1.62 ± 0.03 | 42.7 | 0.965 |
| SMP 7 | 99.1 ± 8.7 | 21.1 ± 0.5 | 1.68 ± 0.03 | 47.2 | 0.894 |
| SMP 8 | 123.0 ± 14.6 | 22.5 ± 1.4 | 1.78 ± 0.05 | 49.9 | 0.902 |
| SMP 9 | 202.0 ± 15.4 | 26.0 ± 0.2 | 1.79 ± 0.03 | 44.2 | 1.176 |

**Table 2**. IEP of colloidal AFM probe and reference flat substrates.

| ***Sample*** | ***$pH_{IEP}$*** |
|---|---|
| Borosilicate glass (colloidal probe annealed at 780°C) | 3.20 ± 0.05 |
| Borosilicate glass (coverslip annealed at 600°C) | 2.82 ± 0.05 |
| $TiO_2$ flat, polycrystalline rutile | 6.28 ± 0.05 |
| $TiO_2$ flat, single-crystal <100> rutile | 3.47 ± 0.05 |

**Table 3**. IEP of ns-$TiO_2$ samples.

| ***Sample*** | ***$pH_{IEP}$*** |
|---|---|
| SMP1 ($R_q$ = 5nm) | 5.20 ± 0.05 |
| SMP2 ($R_q$ = 10nm) | 4.30 ± 0.05 |
| SMP3 ($R_q$ = 14nm) | 4.10 ± 0.10 |
| SMP4 ($R_q$ = 17nm) | 3.70 ± 0.04 |
| SMP5 ($R_q$ = 19nm) | 3.20 ± 0.05 |
| SMP6 ($R_q$ = 20nm) | 3.20 ± 0.05 |
| SMP7 ($R_q$ = 21nm) | 3.20 ± 0.05 |
| SMP8 ($R_q$ = 22nm) | 3.20 ± 0.05 |
| SMP9 ($R_q$ = 26nm) | 3.09 ± 0.04 |

**Figure 1**

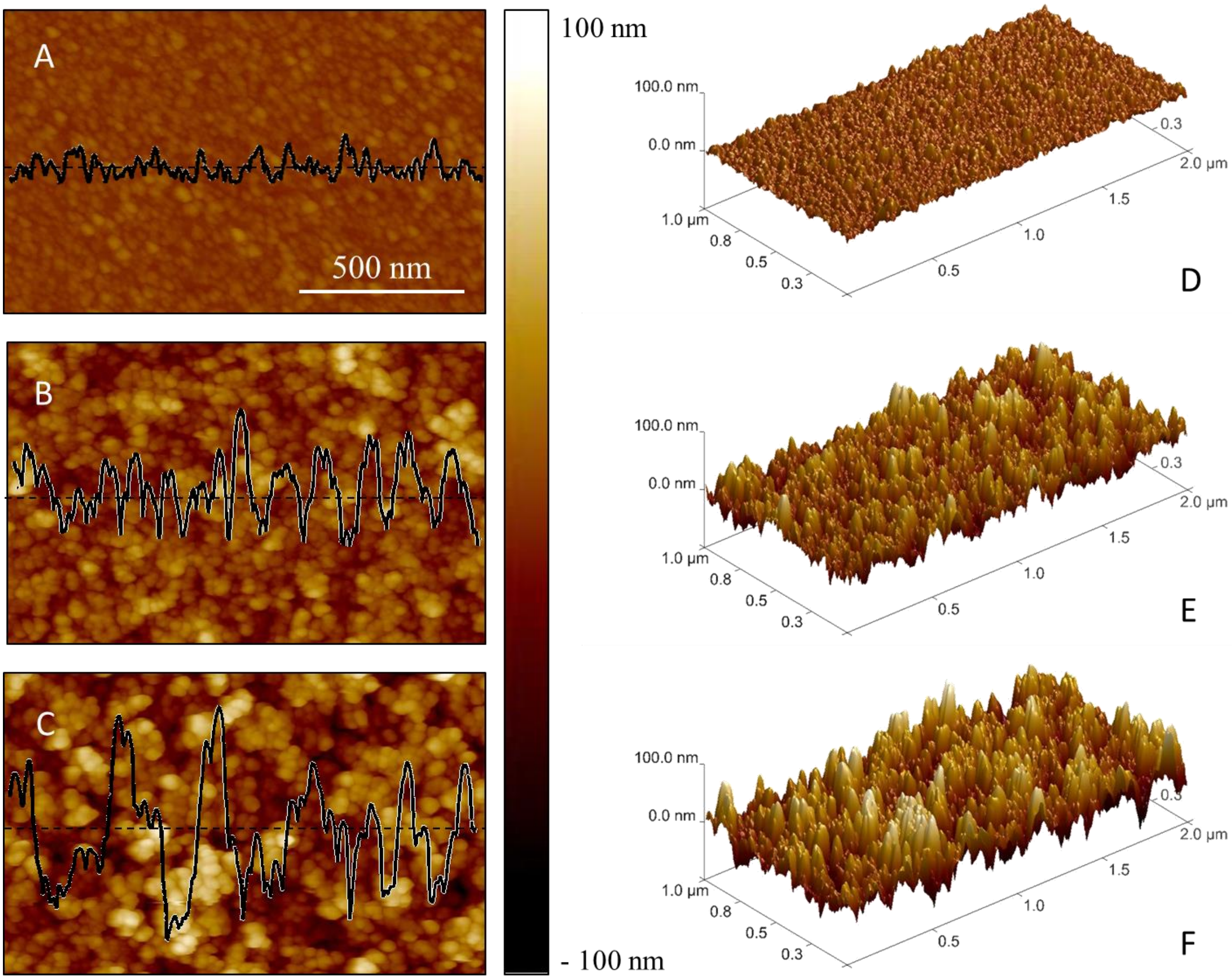

**Figure 2**

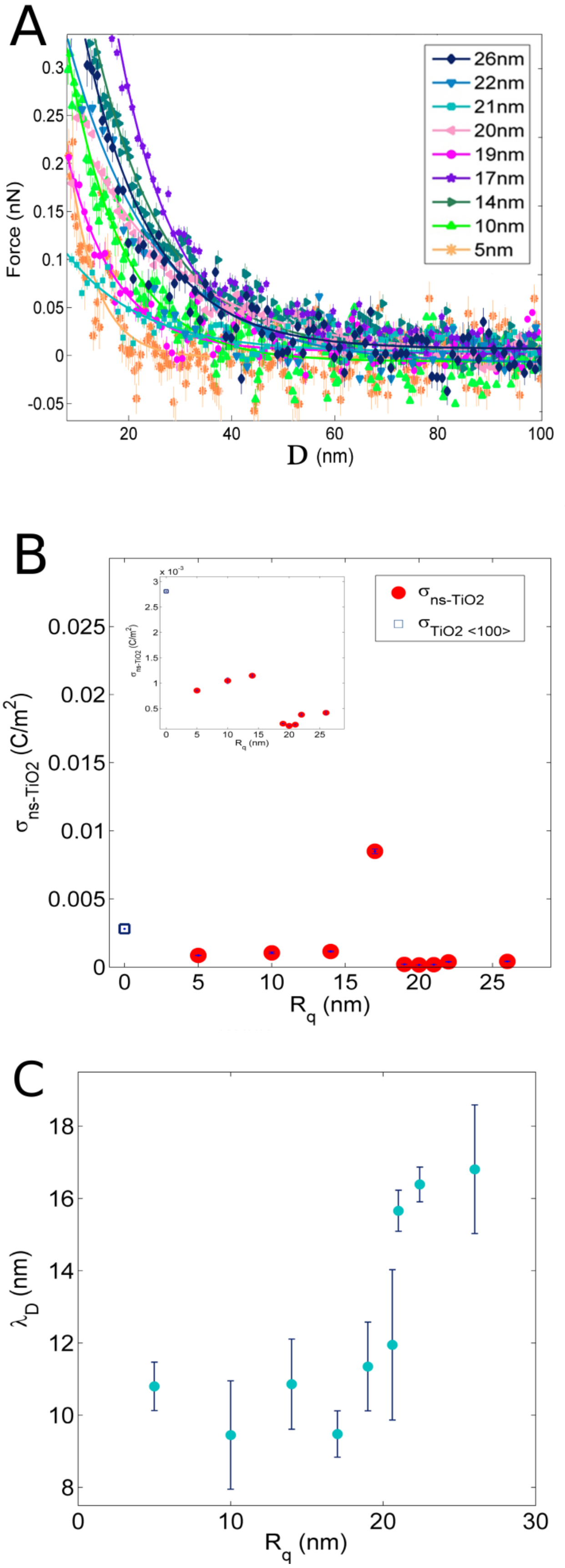

**Figure 3**

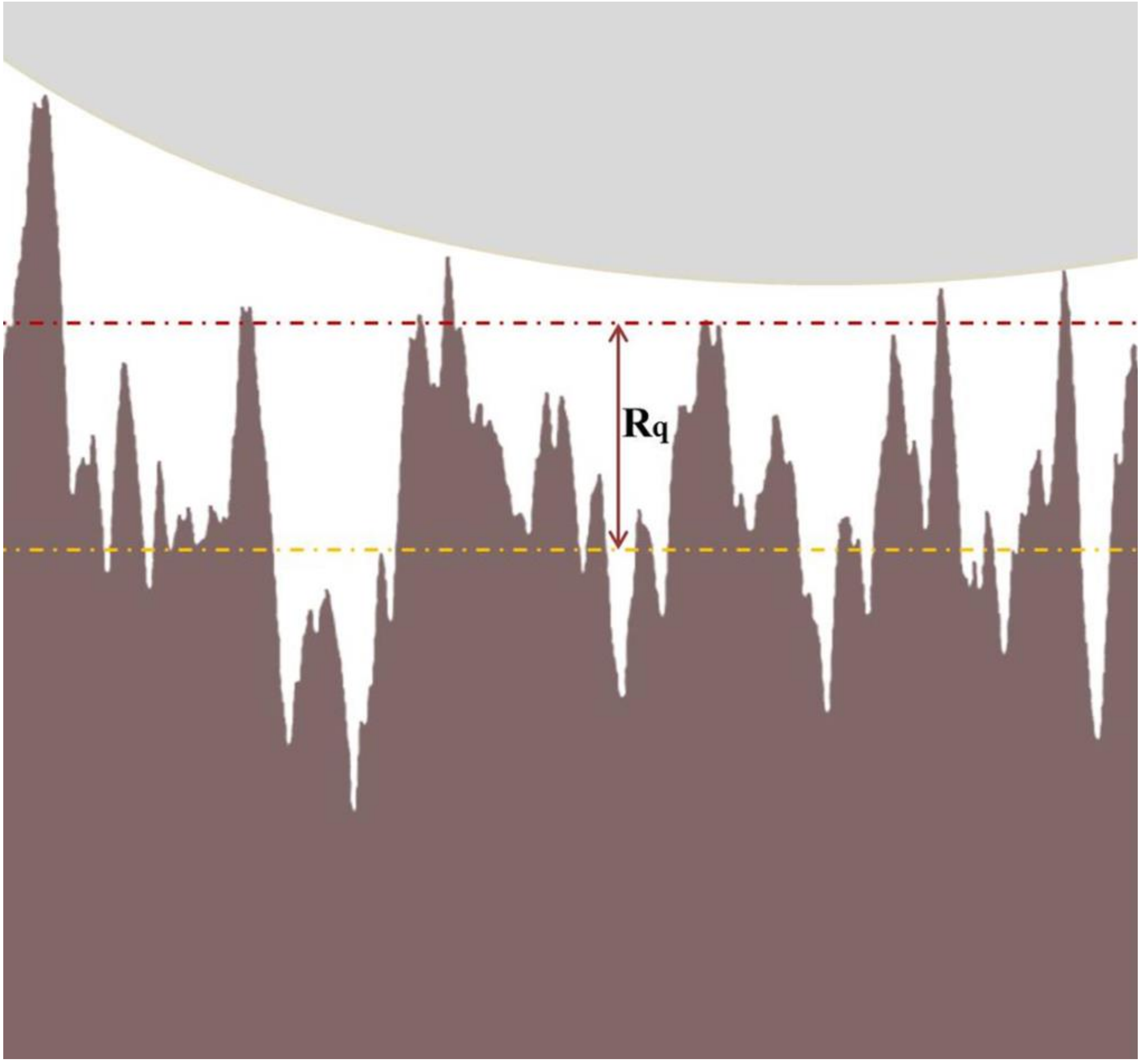

**Figure 4**

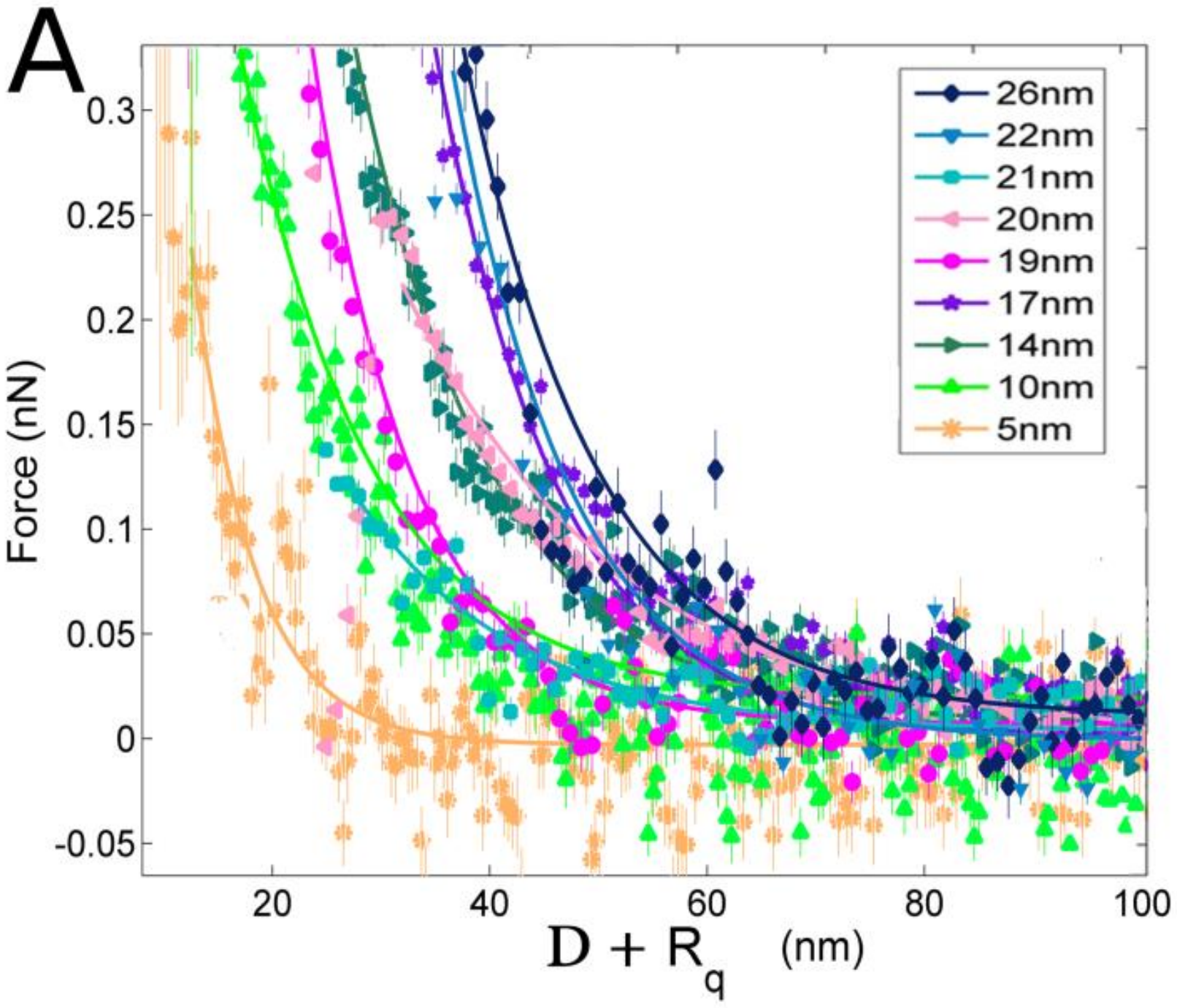

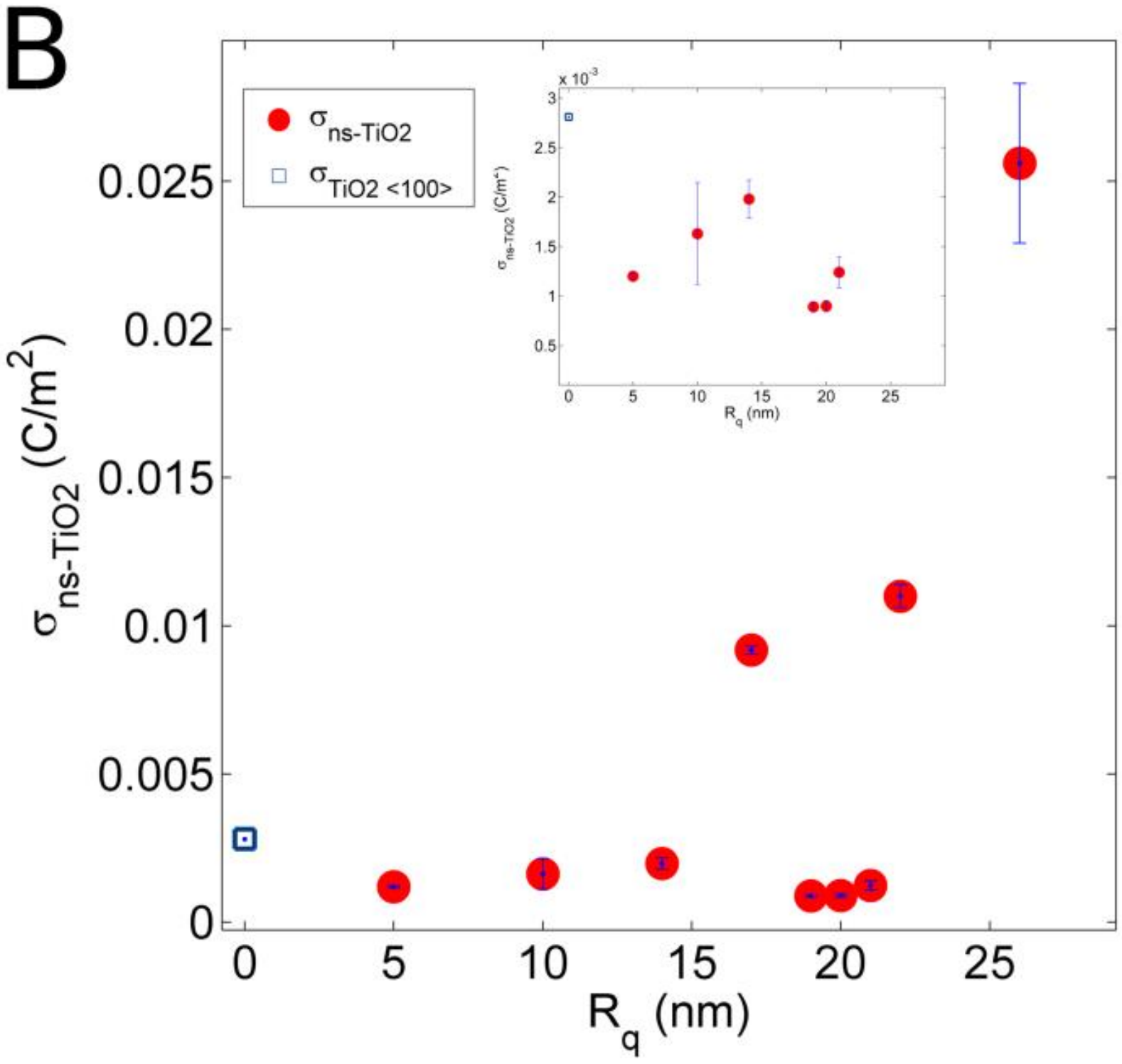

**Figure 5**

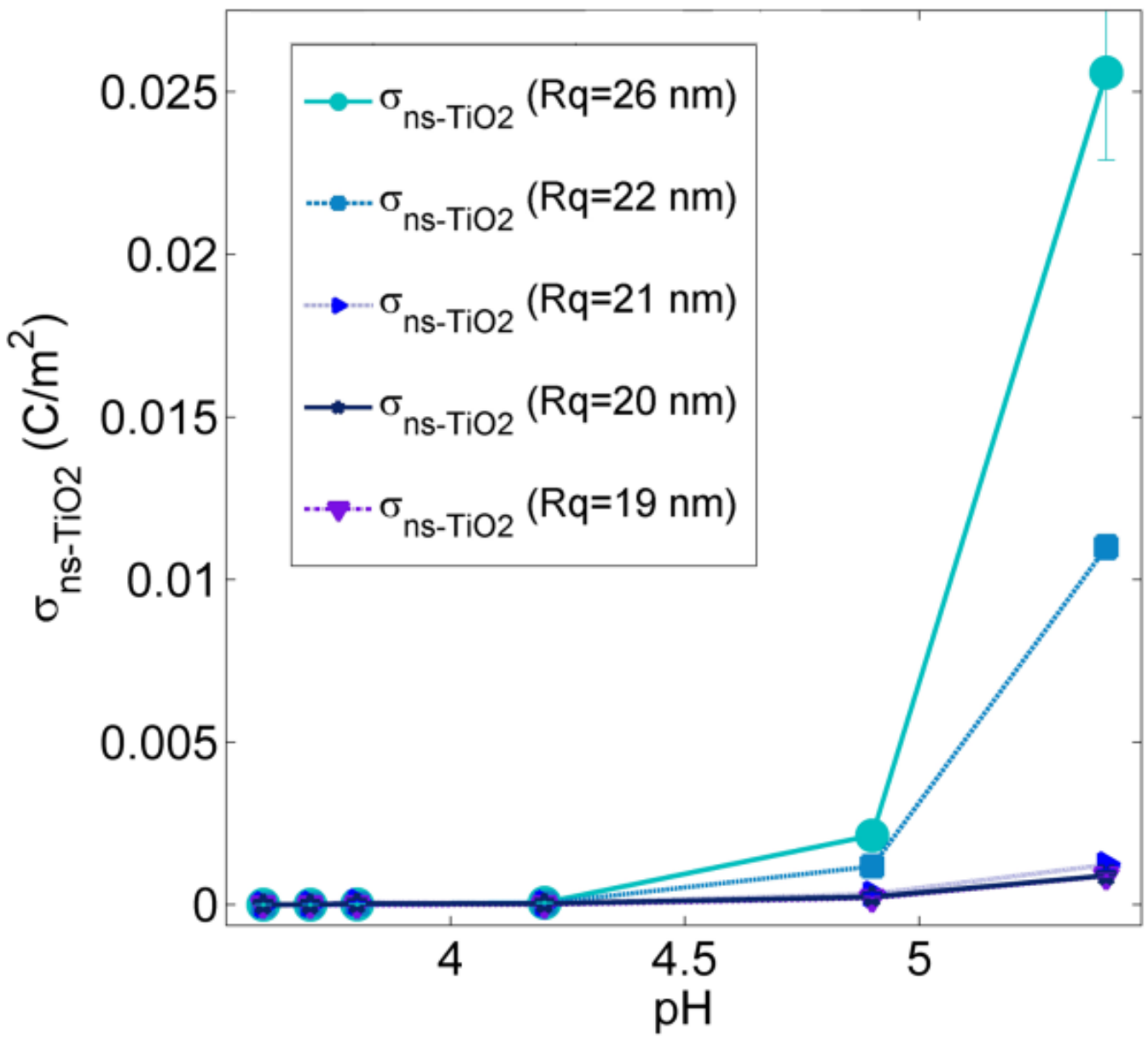

**Figure 6**

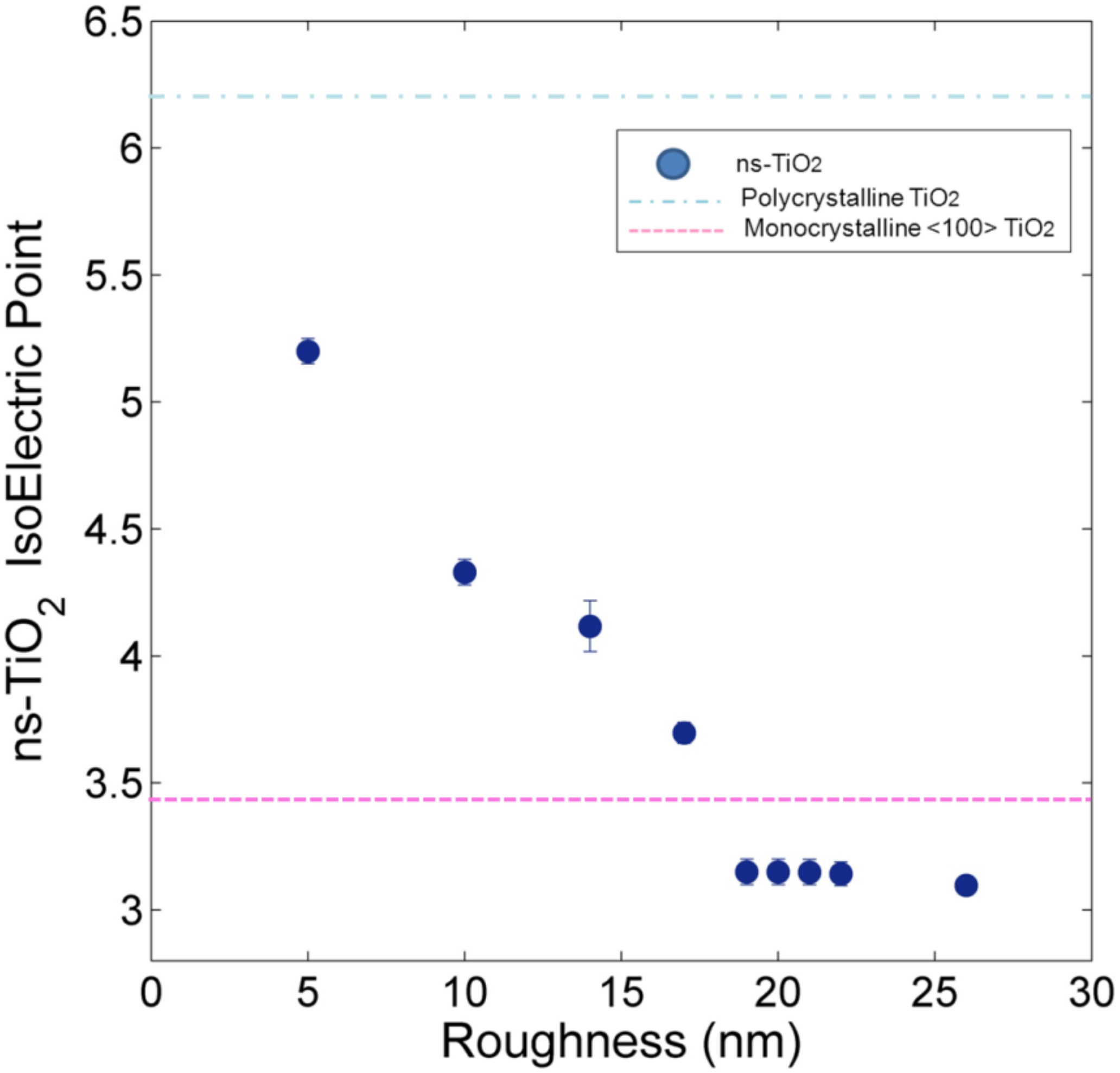

**Figure 7**

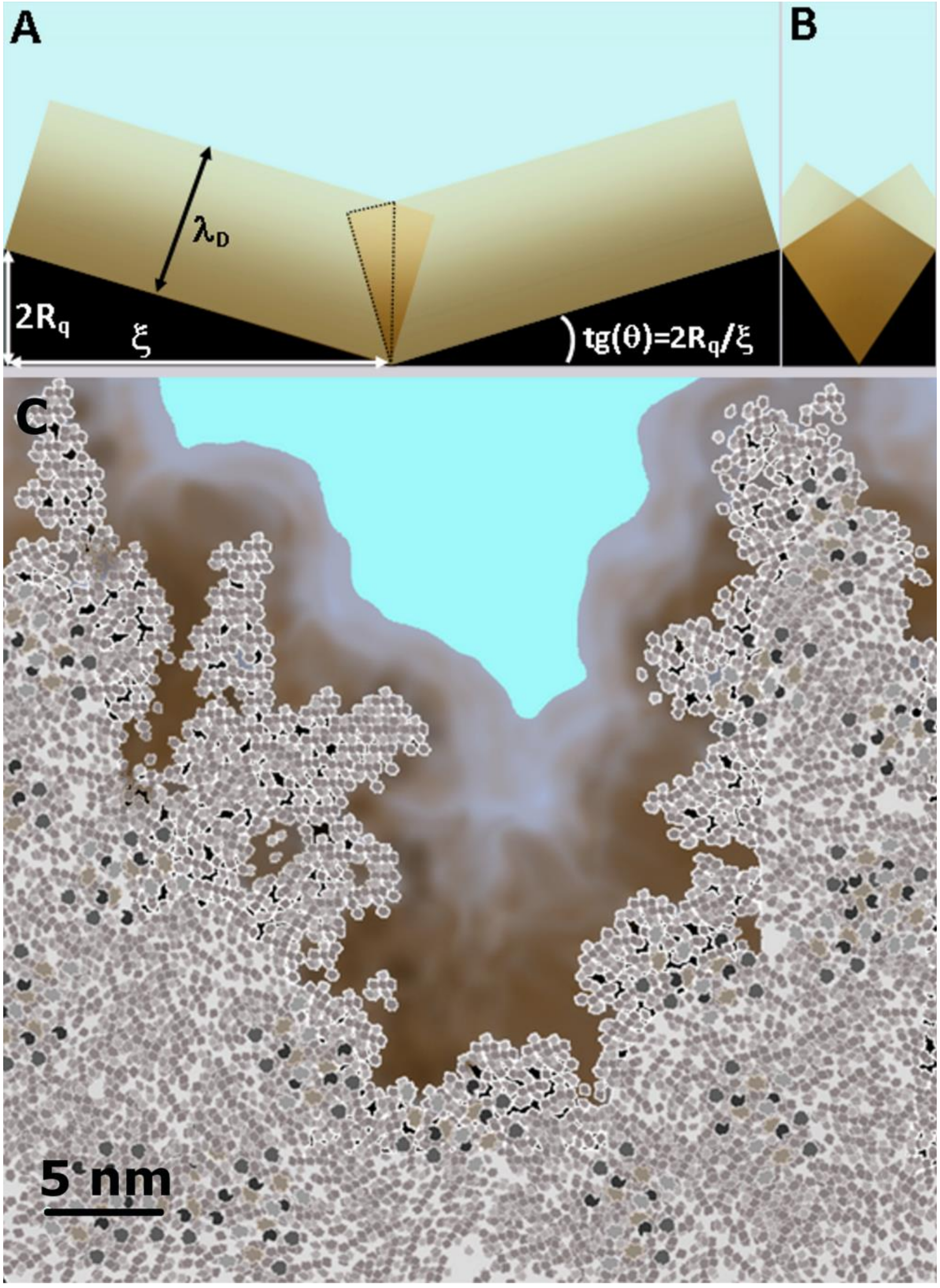

**Nanoscale roughness and morphology affect the IsoElectric Point of titania surfaces**

F. Borghi[1], V. Vyas[1,2,†], A. Podestà[1]*, P. Milani[1]

*1) CIMaINa and Dipartimento di Fisica, Università degli Studi di Milano, via Celoria 16, 20133 Milano, Italy.*

*2) European School of Molecular Medicine (SEMM), IFOM-IEO Campus, Via Adamello 16, 20139 Milano, Italy.*

*† Present address: Institute of Material Sciences, University of Connecticut, 97 North Eagleville Road, Storrs CT, Connecticut, United States.*

* Corresponding author. E-mail: alessandro.podesta@mi.infn.it

# SUPPORTING INFORMATION

## Table of Contents

## 1. Characterization of surface morphology by Atomic Force Microscopy

AFM images were processed using custom routines written in a Matlab environment. The RMS roughness ($R_q$) is calculated as $Rq = \sqrt{\frac{1}{N}\sum_{i,j}(h_{ij} - \bar{h})^2}$, where $h_{ij}$ are height values in the topographic map (i,j are the row and column indices) and N is the number of pixels in the map, $\bar{h}$ is the average height ($\bar{h} = \frac{1}{N}\sum_{i,j} h_{ij}$). The specific area $A_{spec}$ is the ratio of the three-dimensional area calculated on the image to the projected area, i.e. to the AFM scanning area. It is calculated as $A_{spec} = \frac{1}{N}\sqrt{1 + |\nabla h_{ij}|^2}$, where $|\nabla h_{ij}|$ is the modulus of the discretized surface gradient.

The specific area calculated from AFM images is always underestimated because of the inability of the AFM tip to detect overhangs and because of its finite size (typical AFM $A_{spec}$ values do not exceed 2). The in-plane correlations of self- affine surfaces (or profiles) are described by two exponents: the Hurst exponent *H* and the correlation length ξ, which is the characteristic length over which two randomly chosen points on the surface (or on the profile) have uncorrelated heights. The average quadratic difference between heights of two points separated by a distance Δx (also called the height-height correlation function) scales as $\Delta x^{2H}$ for $\Delta x < \xi$, then it saturates. An example is provided in Fig. S1 (here $\sigma \equiv R_q$) and $C_2$ is the h-h correlation function squared.

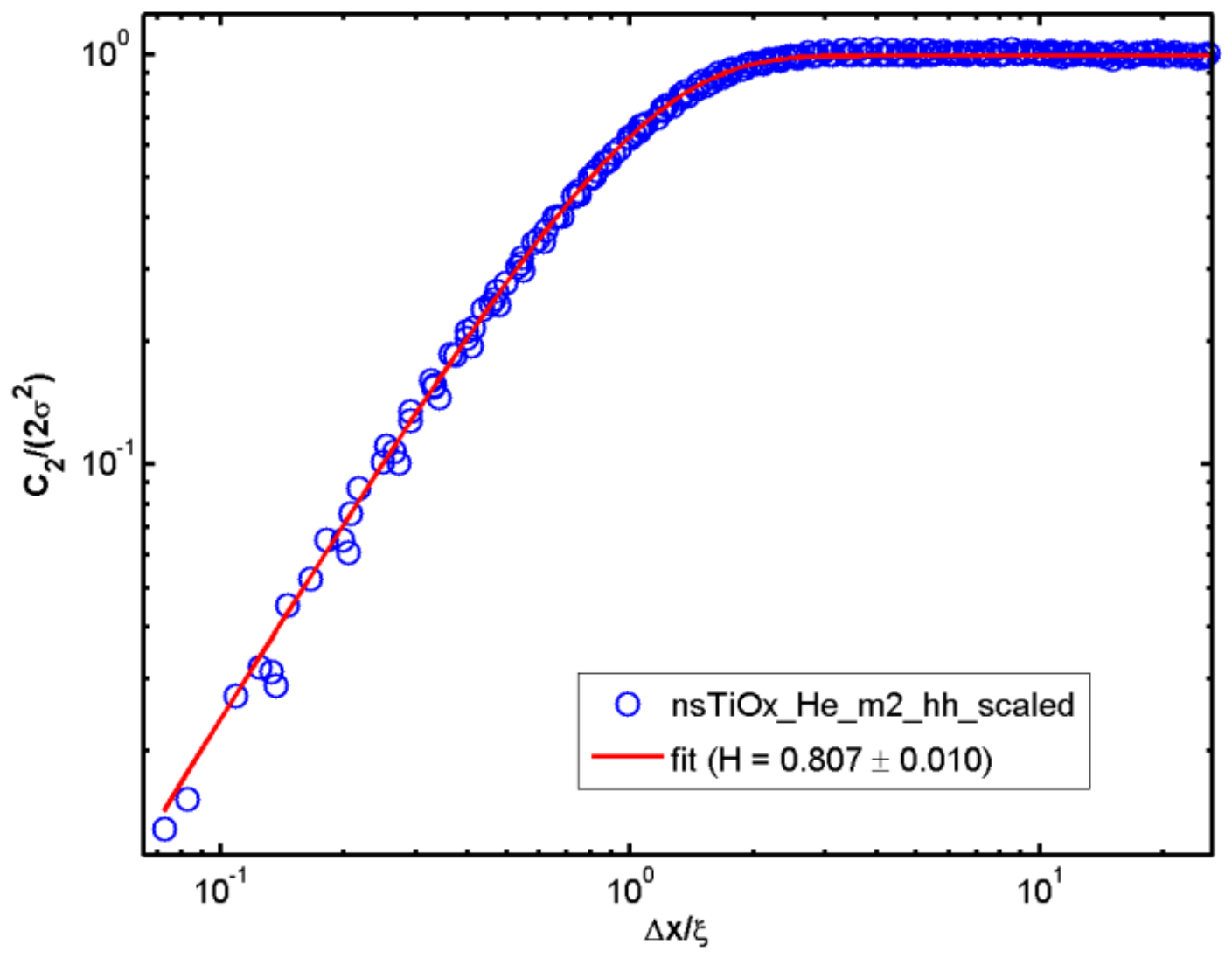

**Figure S1**. Initial linear region and saturation of the height-height correlation function.

The mesoscopic slope of the interface can be calculated as $2R_q/\xi$ (see Fig. 7A in the main text; this result is strictly valid only for a Gaussian surface [1]). For a surface with gaussian distribution of surface heights, the mesoscopic specific area can be calculated as $A_{spec} = 1+2(R_q/\xi)^2$ [1]. Being the determination of both $R_q$ and $\xi$ reliable, the estimation of the mesoscopic specific area is such, as well; it has to be noted that this mesoscopic value fails in reproducing the gain in available area due to sub-correlation length surface structures. Table 1 in the main text reports the value of the punctual specific area calculated directly from AFM topographical maps as described above.

## 2. Characterization of colloidal probe radius

We have calibrated the radius of colloidal probes following a procedure recently introduced [2]. We have imaged a calibration grating array (MikroMasch TGT01) of sharp spikes with apical radius less than 10nm and tip angles below 25° with a cantilever equipped with the colloidal probe used in the DLVO experiments. The lateral and diagonal separations of the spikes are 2.12 and 3 μm, accordingly, and the height of the spikes is in the 600-800 nm range. Therefore, due to the very high aspect-ratio of reference sample features convolution between the geometry of the colloidal probe and the sample morphology is at its maximum, and the captured image (Fig. S2A) is the inverted image of the AFM probe, which can be modeled by spherical caps (Fig. S2B).

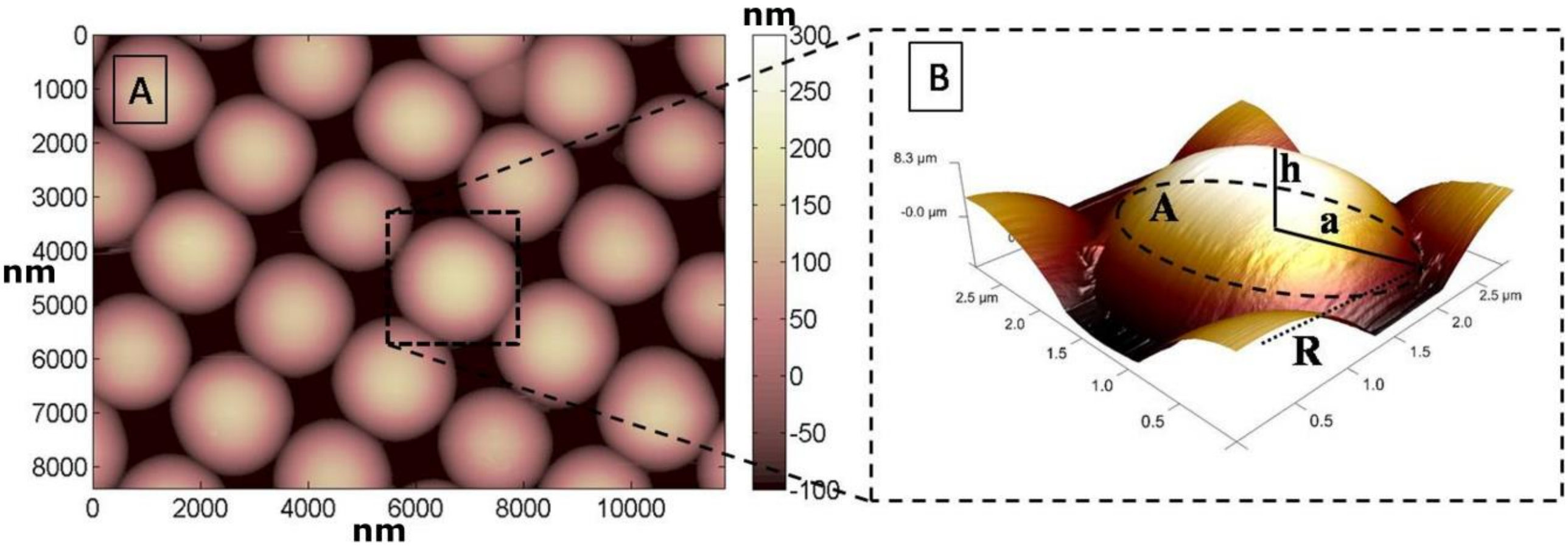

**Figure S2**. (A) AFM image of TGT01 grating obtained with a colloidal probe of nominal radius R=2.5 μm. Axes are in nm units. (B) 3D magnified view of an inverted AFM image of the probe and its geometrical parameters: the base radius *a* and area $A=\pi a^2$, the height h of the spherical cap and the radius of the mother sphere R.

The value of the probe radius and its error are extracted applying a statistical analysis of AFM topographs. In Fig. S3 are shown the Volume vs Height data of the spheres and the estimation of the radius extracted from the fit to the equation $V = \pi/3\ h^2\ (3R - h)$, V and h being the measured volume and height of the spherical caps found in AFM topographic maps like the one shown in Fig.

S2A ($R_{best}$= 2170 ± 65 nm). The microsphere has been attached to a rectangular tipless silicon cantilever (Nanosensors) and the force constant, determined by thermal tuning [3], is 0.43 N/m.

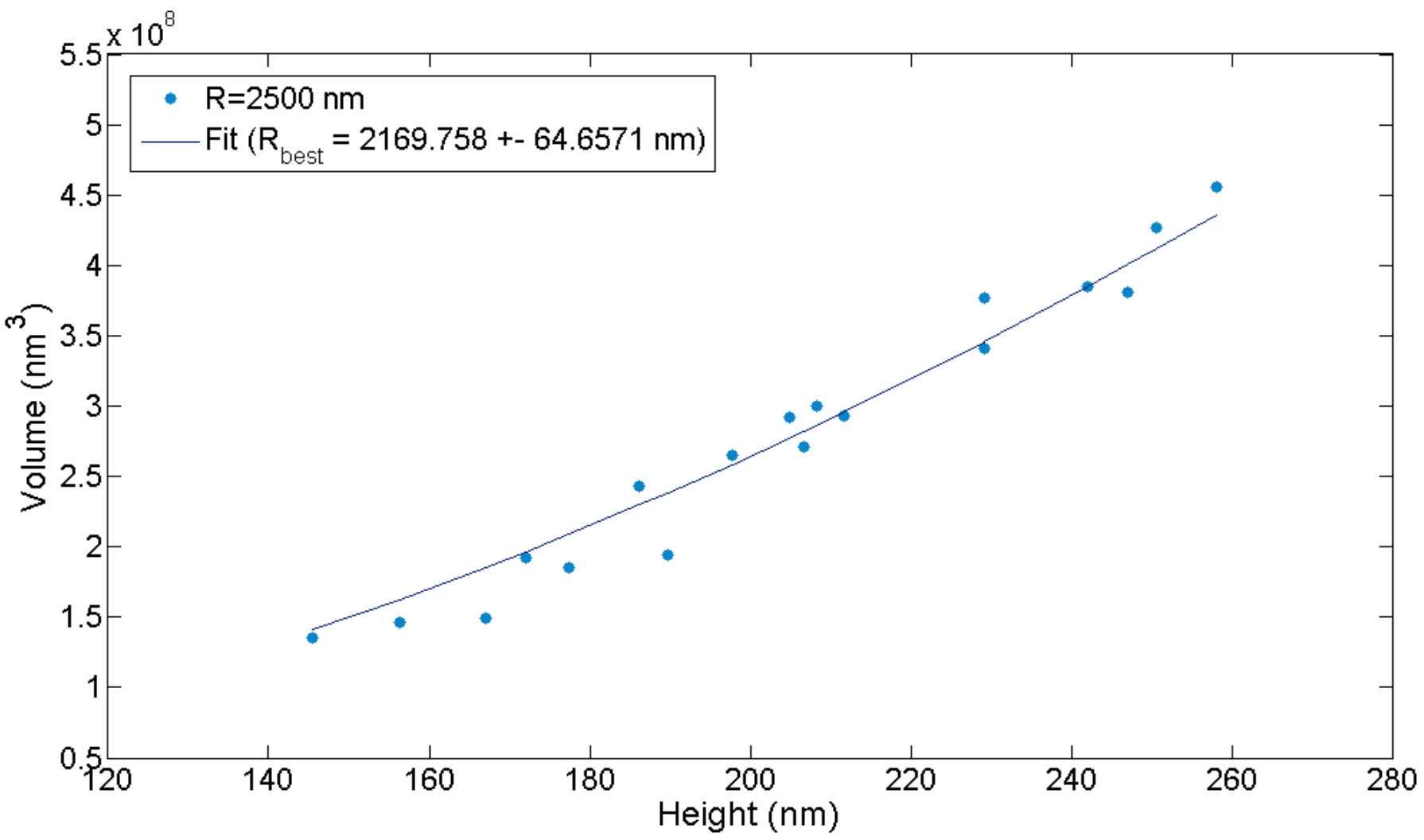

**Figure S3.** Volume versus height data extracted by the inverted AFM images of the colloidal probe used in the DLVO experiments on the TGT01 calibration sample.

## 3. Details on force curves and curve fitting procedures

Typical average force curves, acquired using a borosilicate glass colloidal probe with radius R=2170±65 nm on a flat glass borosilicate surface with [NaCl] varying in the range 0.1-100 mM and fixed neutral pH (≈6.5), are shown in Fig. S4.

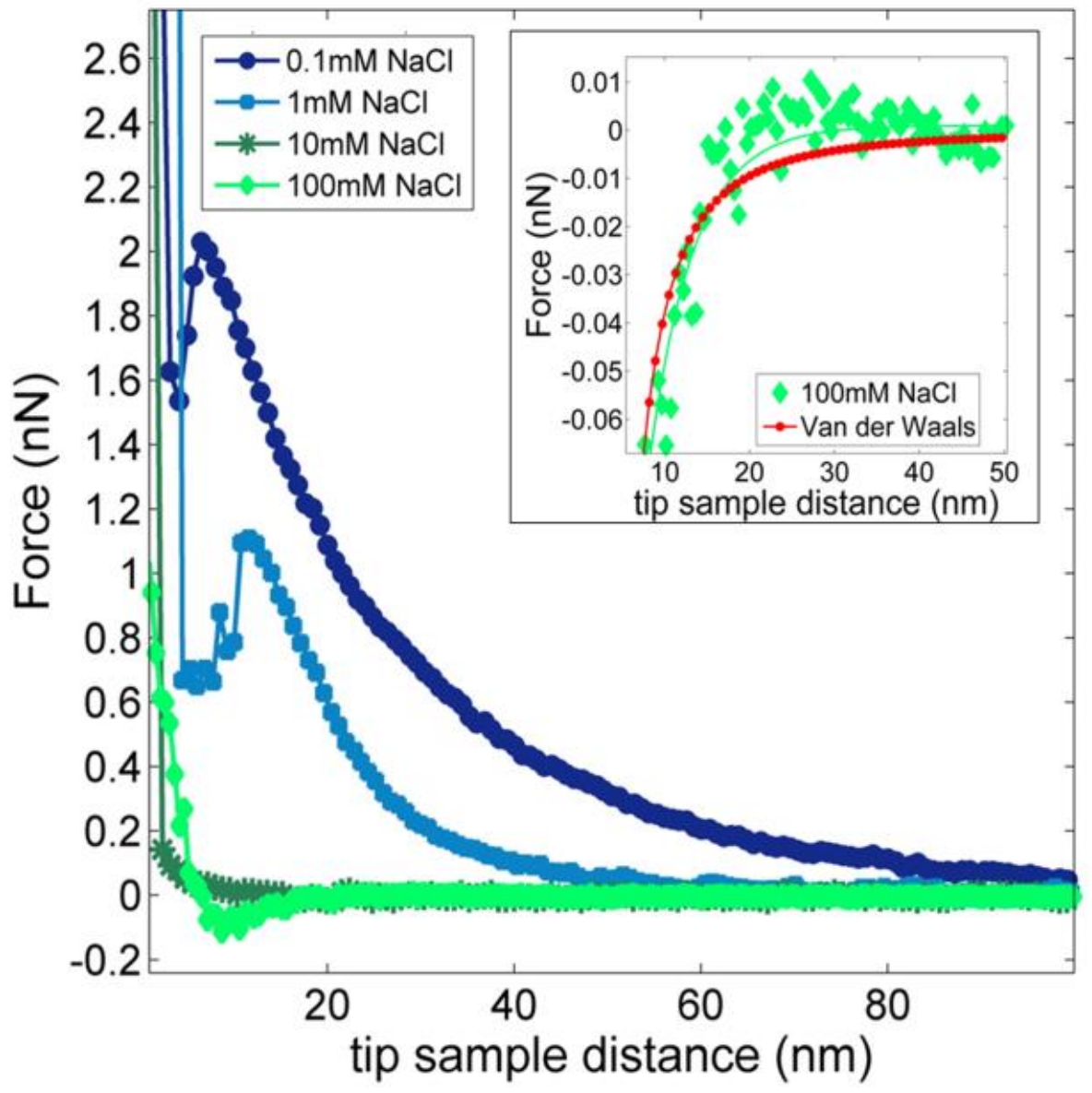

**Figure S4.** Average force curves between borosilicate glass colloidal probe and a flat borosilicate glass coverslip, acquired in solution with different ionic strength (0.1mM – 100mM NaCl). In the inset it is shown the overlapping between Van der Waals force curve (calculated using A=0.8*$10^{-20}$ J) and the experimental curves in 100mM NaCl solution.

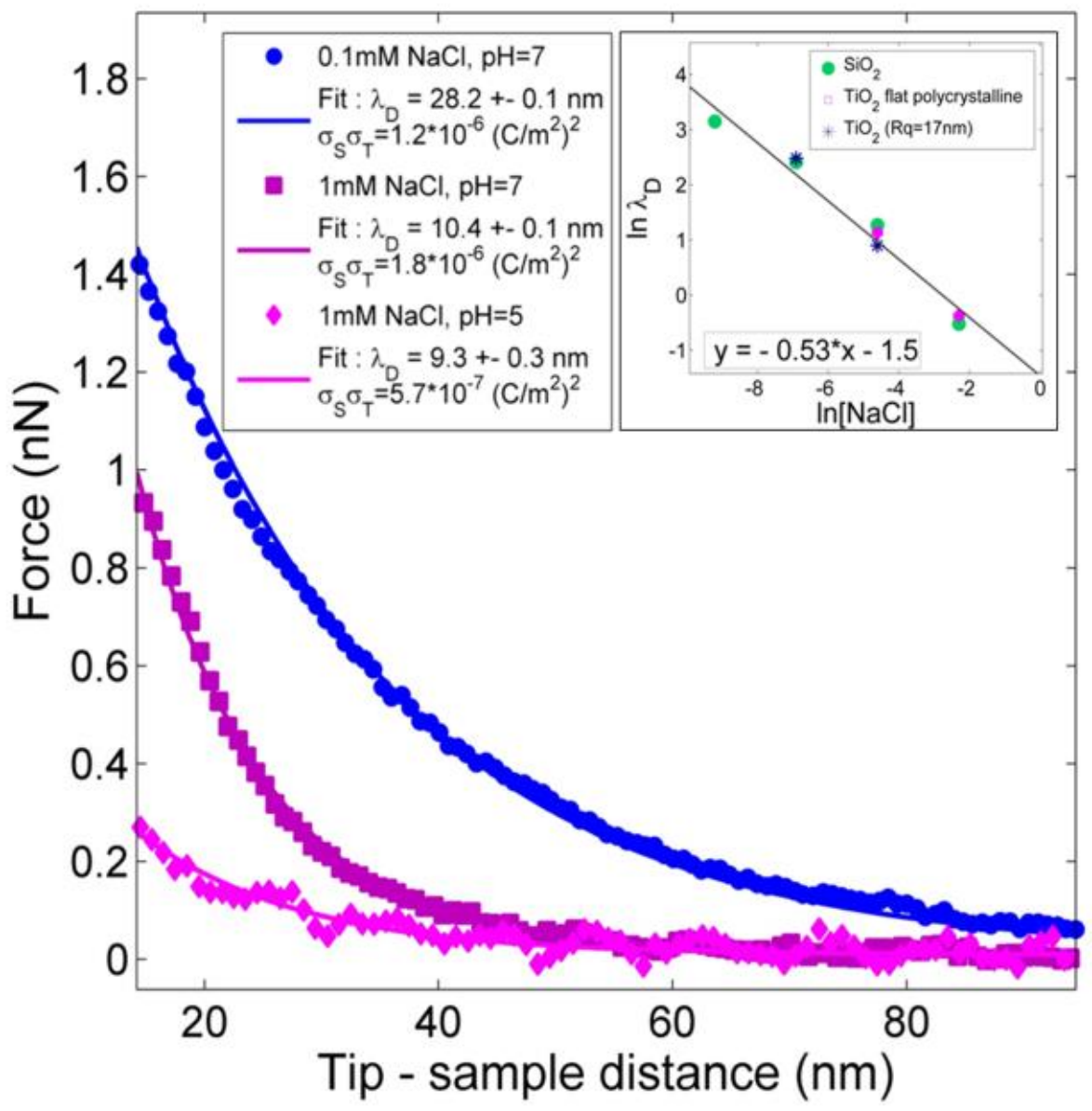

**Figure S5.** Best fit of average force curves between borosilicate glass colloidal probe and a flat borosilicate glass coverslip, acquired in solution with different ionic strength (0.1mM and 1mM NaCl, pH=7) or with the same ionic strength but different pH (1mM NaCl, $pH_1$=7 and $pH_2$ =5). In the inset, log plot of the Debye length versus the inverse of the square root of NaCl concentration, calculated in experiments with different substrates.

Error bars on average force data (not shown here, see main text) are calculated summing in quadrature two errors: a statistical error, typically negligible, calculated as the standard deviation of the mean of force values that are averaged, and a systematic error due to the calibration of the AFM cantilever, which is determined considering a 2% error due to deflection sensitivity calibration (see Mats&Methods in the main text) and 5% error due to the force constant calibration. Interpretation of force curves is the following. The tip, approaching the surface, remains in its rest position (constant deflection signal) until at a certain distance from the surface, depending on the ionic strength of the solution, it feels first the long-range electrostatic interaction with the sample surface and subsequently the Van der Waals attraction force [4,5]. An increased salt concentration (or an increased Ionic strength of the solution) determines a decrease of the electrostatic force, even if the repulsion grows steeper. At the same time, the jump-in due to the Van der Waals attraction, takes place at larger distance from the surface; by increasing the salt concentration it shifts from 7 to 18nm. The smearing of the force curves at short distances is an artifact caused by the averaging process, due to the fact that the jump-in distance fluctuates by several nm from curve to curve; DLVO fit is performed in the large-distance region, typically between 10 and 100 nm, well before the onset of the jump-in. At the highest salt concentration the electrostatic repulsion is completely overwhelmed by Van der Waals attraction; a minimum appears, due to van der Waals force, while only at the shortest distance electrostatic repulsion can be appreciated. An expanded view of this curve is shown in the inset of Fig. S4, together with the Van der Waals contribution evaluated by the second term of Eq. 6 (main text) using A=0.8 $10^{-20}$ J.

It is very important to control the pH of the solution before and after AFM measurements in order to check the stability of the system and guarantee the accuracy in the determination of the IEP. It is also important to wait more than fifteen minutes after the immersion of the thin film and tip in the solution and to rinse the surfaces, before and after measurements, with neutral distilled water, in order to reach the equilibrium stability and to restore surface charges. Experimental data confirm that different ionic strengths determine the value of the Debye length according to Eqs. 3,4 without affecting the $\sigma_S\sigma_T$ value, while for the same value of Ionic Strength, $\sigma_S\sigma_T$ decreases with the pH of the solution until the value equals the first IEP of the system. Representative force curves and their best fit (using Eq.6) are shown in Fig. S5. The inset of Fig. S5 shows experimental values of $\lambda_D$ measured in different salt concentrations solution, with different surfaces ($SiO_2$, flat polycrystalline $TiO_2$ and rough ns-$TiO_2$). $\lambda_D$ scales as the inverse of the square root of $[NaCl]^{-1/2}$, as predicted by Eq. 4. We have also verified the stability of the solutions characterized by different value of pH during a period of one month, in the pH range between 3 and 7. pH values were checked using a pH meter. We have chosen to fix the 1mM NaCl concentration because it allows us to analyze a large range of pH values without changing Ionic strength of the solution and also because, in a more concentrated solution, the 1:1 electrolyte is no more completely inert for $SiO_2$ and $TiO_2$, promoting a shift of the IEP. Furthermore, for 1mM NaCl solution, the Debye length ($\lambda_D$ ~ 9.6 nm) is large enough to guarantee a wide interval of electrostatic interaction and a higher signal-to-noise ratio.

| **Borosilicate glass probe vs borosilicate glass substrate** | **Borosilicate glass probe vs rutile 100 $TiO_2$** | **Borosilicate glass probe vs ns-$TiO_2$ (pH=5.4)** |
|---|---|---|
| pH=3.4 | pH=4.28 | Rq=14 nm |
| 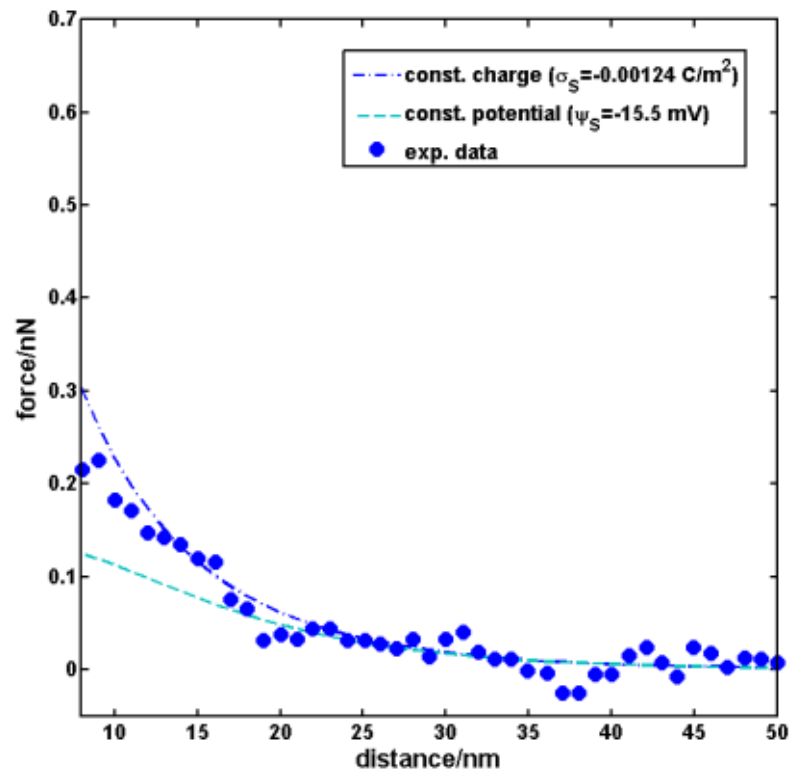 | 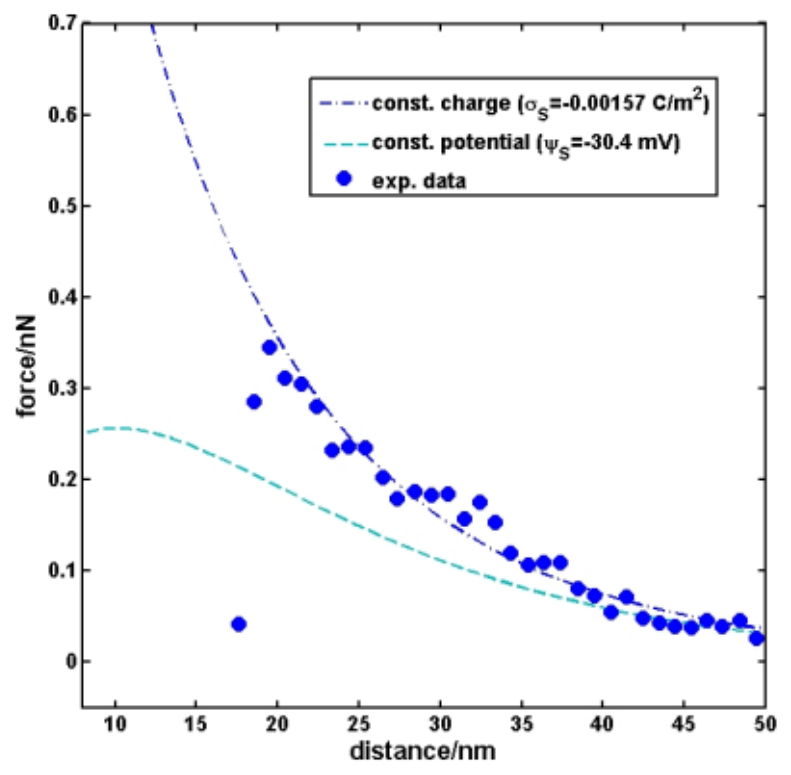 | 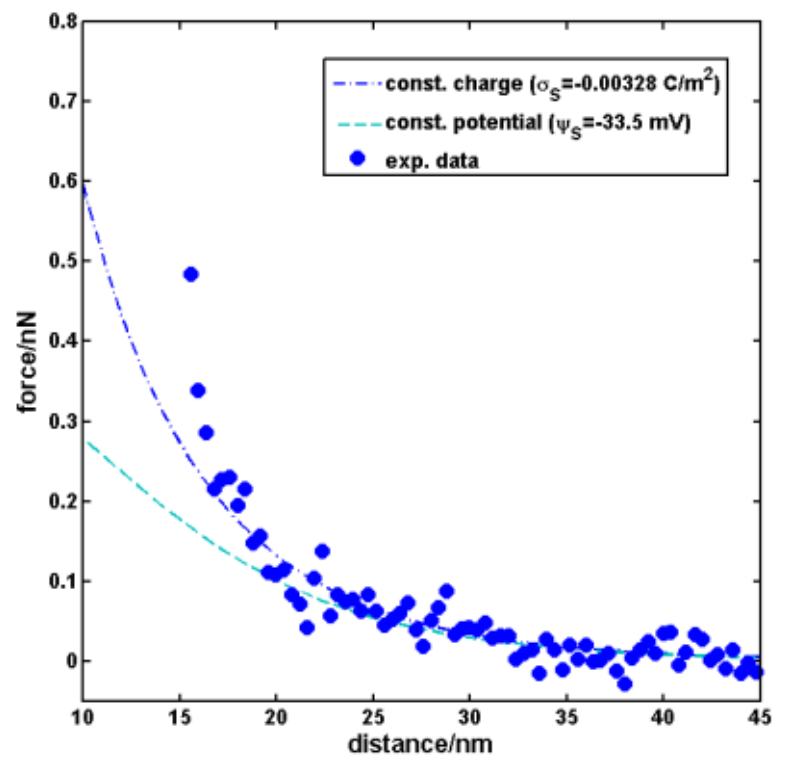 |
| pH=5.25 | pH=5.4 | Rq=22 nm |
| 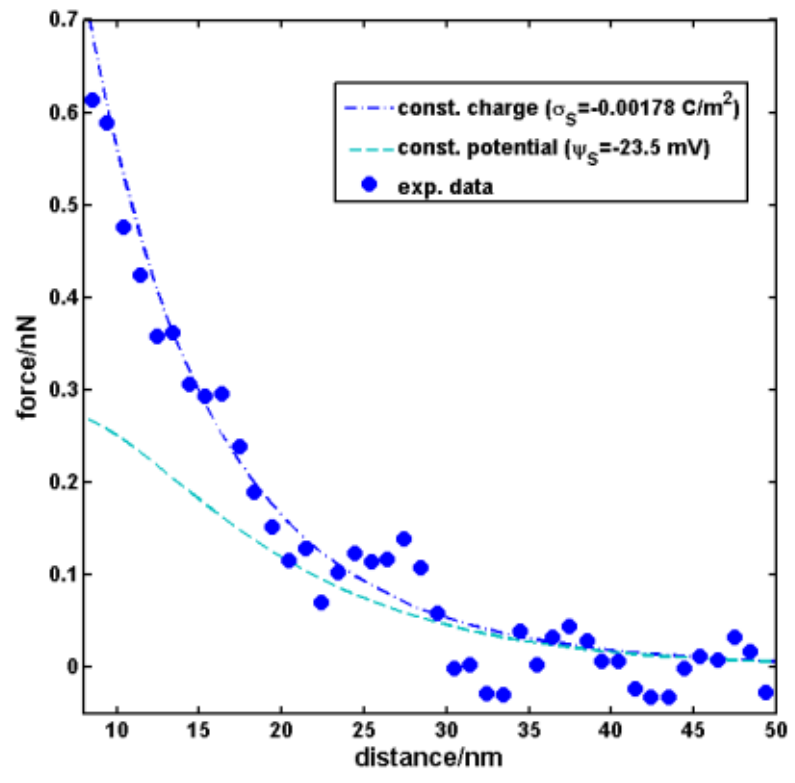 | 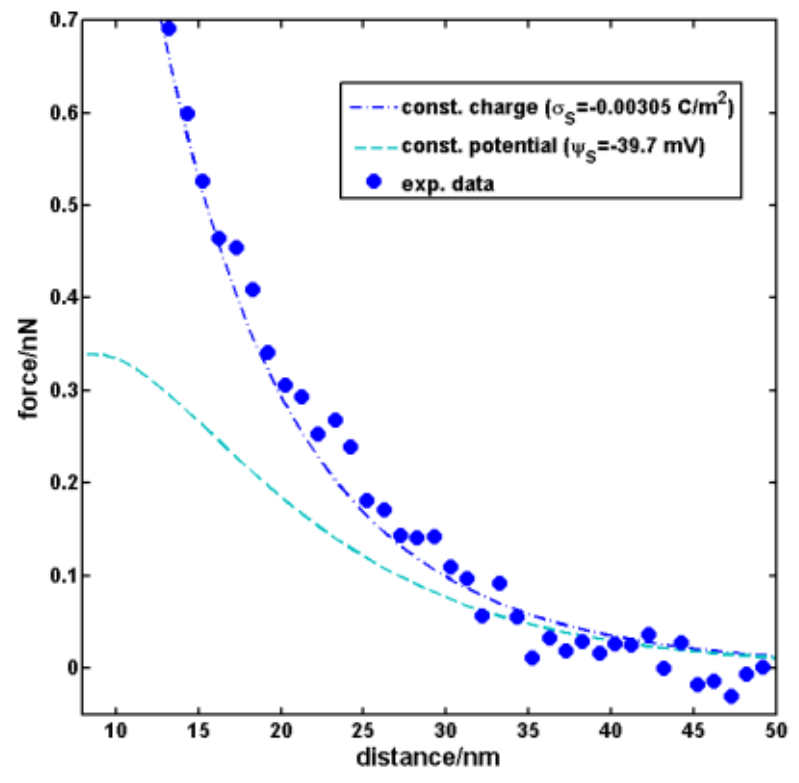 | 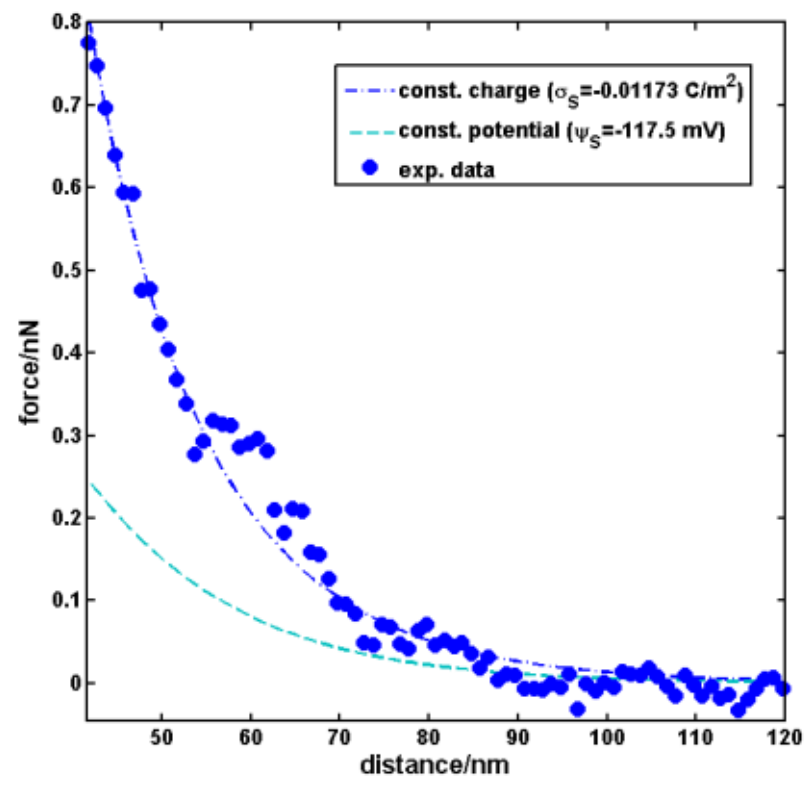 |
| pH=6.8 | pH=7.9 | Rq=26nm |
| 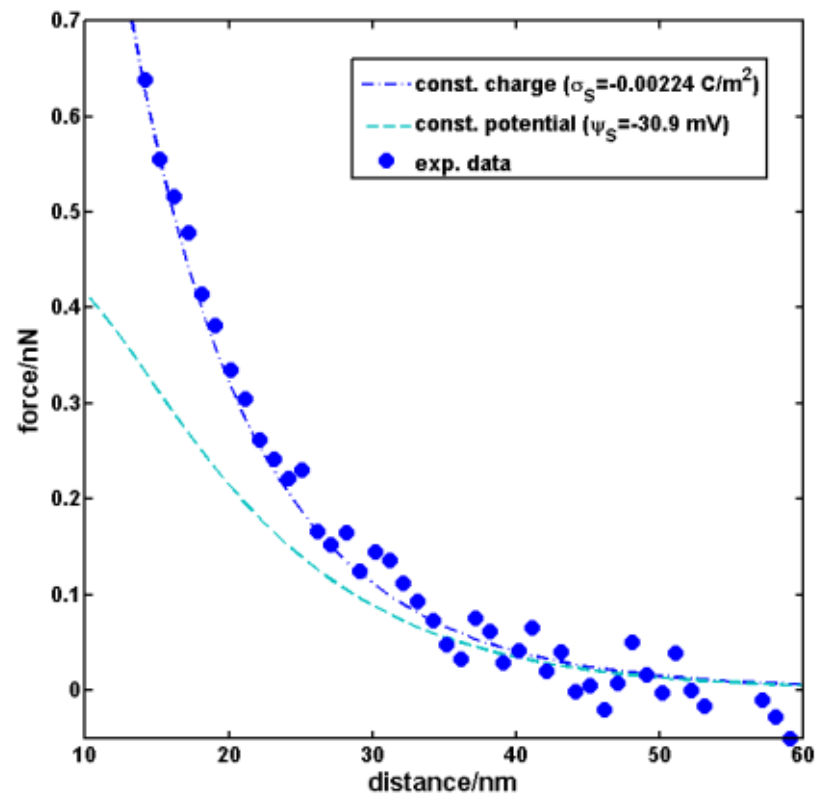 | 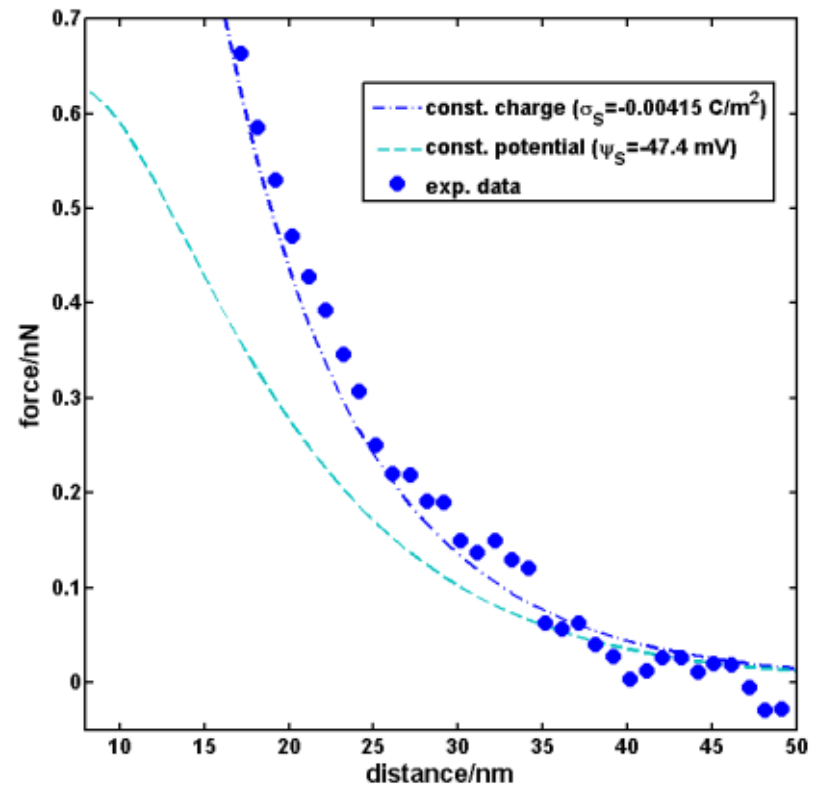 | 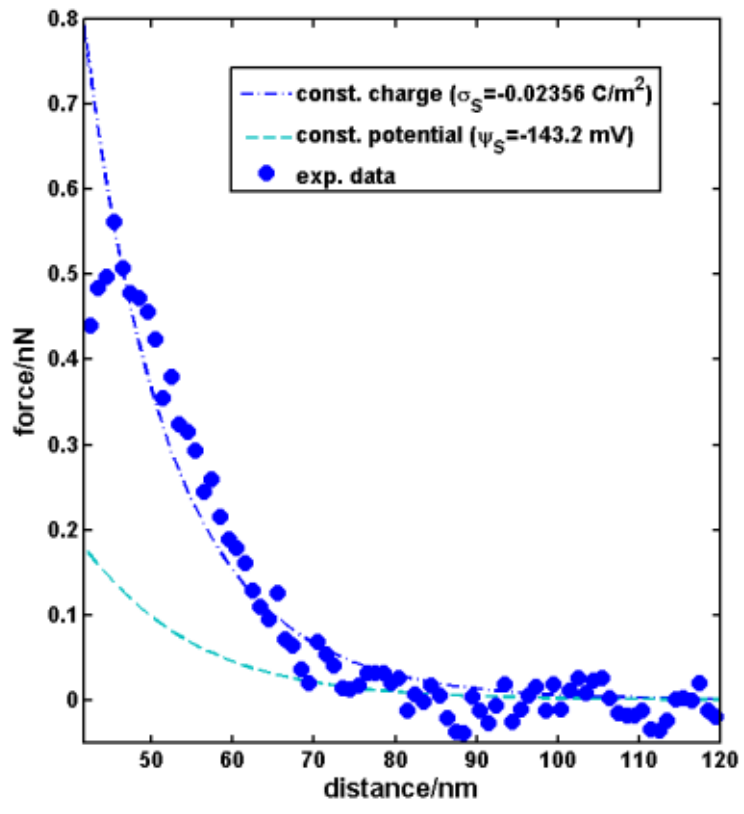 |

**Figure S6.** Comparison of force data acquired using a borosilicate glass colloidal probe on borosilicate glass substrate, rutile <100> and nanostructured $TiO_2$, with constant charge and constant potential curve obtained from nonlinear regression via Eq. 1 and from Eq. 2, using potentials calculated by Grahame equation (Eq. 5).

### 3.1. Applicability of the constant charge model for DLVO force

We have tested the applicability of the constant charge DLVO force model (Eq. 1 in the text, and its approximation for larger distances, Eq. 6), which is typically found to describe appropriately DLVO interactions between insulating oxide surfaces in aqueous electrolytes. Both constant charge and constant potential models (Eq. 1 and Eq. 2) overlap at distances sufficiently larger than $\lambda_D$, where Eqs. 1,2 reduce to a single exponential term whose prefactor contains the product of surface charges or surface potential, depending on the boundary conditions; charge densities and potentials are related by Grahame equation (Eq. 5). We have fitted the force curves by Eq. 1 across a distance range exceeding $1.5\lambda_D$, and used Grahame equation to calculate the diffuse layer potentials from the values of the diffuse charge densities $\sigma_S$ and $\sigma_T$ (the AFM probe-borosilicate glass substrate system was considered symmetric, which allowed to determine the absolute charge density of the probe; the latter parameter was kept fixed in fitting curves of other systems). It turned out that constant potential force curves systematically underestimate experimental data (Fig. S6), while the constant charge model could fit data across the complete range of distances (from jumpin to about 50 nm).

### 3.2.Fitting strategy

Eq. 1 and Eq. 6 overlap at sufficiently large distances; by fitting the force curves data with Eq.6 at distances larger than approximately $1.5\lambda_D$ it was possible to determine the Debye length and the product $\sigma_S\sigma_T$ of charge densities. If one knows the charge density of one of the two surfaces, the other can be determined. In particular, on symmetric systems $\sigma_S \approx \sigma_T$ and therefore $\sigma_T \approx \sqrt{\sigma_S\sigma_T}$ . We could therefore characterized the net surface charge density of the colloidal probe from force measurements in aqueous electrolyte on a borosilicate glass substrate (see section 3.1 of Supporting Information); we have then used the values of $\sigma_T$ at different pH to calculate the absolute net charge density $\sigma_S$ of crystalline and nanostructured $TiO_2$ surfaces.

For each sample 100 force curves were typically acquired in six different locations (separated by 100µm) in order to accurately characterize the Debye length and the charge densities of the surfaces. Charge densities and Debye lengths extracted from average force curves of different locations were averaged; their errors were estimated as the 68% confidence interval according to the optimized strategy discussed by Lybanon [6], consisting in repeating the fit on a set of artificial experimental data obtained by summing a Gaussian error to the original data based on errors on both force and distances, then looking at the dispersion of fit parameters obtained. For both Debye lengths and charge densities, the error δ associated to the averages across different locations was calculated propagating the errors $\delta_i$ of the nonlinear regression through the arithmetic mean function, i.e. $\delta = {}^{1}/_{N} \sqrt{\sum_i^N \delta_i^2}$.

## 4. Charging of surfaces in liquid electrolytes

The charging behaviour of metal oxide surfaces in aqueous electrolytes is generally attributed to amphoteric character of surface hydroxyl groups [7-12]. Charging of the solid surface can be formally regarded as a two-step protonation of surface $M\text{-}O^-$ groups:

$$M\text{-}O^- + H^+ \leftrightarrow M\text{-}OH;\ K_1 \quad \text{(S1a)}$$

$$M\text{-}OH + H^+ \leftrightarrow M\text{-}OH_2^+;\ K_2 \quad \text{(S2)}$$

or to the interaction of surface hydroxyls M-OH with $OH^-$ and $H^+$ ions, in which case the first reaction must be replaced with:

$$M\text{-}OH + OH^- \leftrightarrow M\text{-}O^- + H_2O;\ K_1' \quad \text{(S1b)}$$

The equilibrium constants $K_1$ and $K_2$ are defined as: $K_1=[M\text{-}OH]/([M\text{-}O^-][H^+])$ and $K_2=[M\text{-}OH_2^+]/([M\text{-}OH][H^+])$, [X] representing the molar concentration of the species X. It turns out that $1/K_1'=K_wK_1$, $K_w=10^{-14}$ being the equilibrium constant of the dissociation reaction of water into $H^+$ and $OH^-$ ions (due to its very small value, $pK_1$ and $pK_1'$ are almost equal, being $pK=-\log_{10}(K)$).

In addition to association/dissociation of surface hydroxyls described by Eqs. S1,S2, also adsorption of anions $A^-$ and cations $C^+$ from solution to charged surface sites may take place, according to reactions:

$$M\text{-}O^- + C^+ \leftrightarrow M\text{-}O^-\cdot C^+ \quad \text{(S3)}$$

$$M\text{-}OH_2^+ + A^- \leftrightarrow M\text{-}OH_2^+\cdot A^- \quad \text{(S4)}$$

where $K_+=[M\text{-}O^-\cdot C^+]/[M\text{-}O^-][C^+]$ and $K_-=[M\text{-}OH_2^+\cdot A^-]/[M\text{-}OH_2^+][A^-]$.

The surface charge density $\sigma_0$, the charge density at the inner Helmholtz plane $\sigma_I$, and the charge density of the diffuse layer at the outer Helmholtz plane $\sigma_d$ are equal to:

$$\sigma_0 = F\,([M\text{-}OH_2^+] + [M\text{-}OH_2^+\cdot A^-] - [M\text{-}O^-] - [M\text{-}O^-\cdot C^+])$$

$$\sigma_i = F\,([M\text{-}O^-\cdot C^+] - [M\text{-}OH_2^+\cdot A^-])$$

$\sigma_d = -(\sigma_0+\sigma_i) = -F([M\text{-}OH_2^+] - [M\text{-}O^-])$

where F is the Faraday constant, i.e. the number of coulombs per mole of electrons.

## 5. Determination of charge density products and IEPs of reference systems

Force curves have been acquired at 20°C in 1mM NaCl solutions at different pH (from 3 to 8), whose value is detected immediately before and after the AFM measurements by a pH meter. We have fitted the average curves with Eq. 6, for distances larger than approximately 15-20 nm, and sufficiently far away from the jump-in point, in order to avoid the mix-up between electrostatic force and the repulsion in contact regime and to neglect the contribution of the term in Eq. 1 proportional to *exp(-2D/$\lambda_D$).*

### 5.1. Borosilicate glass colloidal probe and reference substrate

In Fig. S7A electrostatic interactions at different pH between the colloidal borosilicate glass tip and the borosilicate glass coverslip are shown.

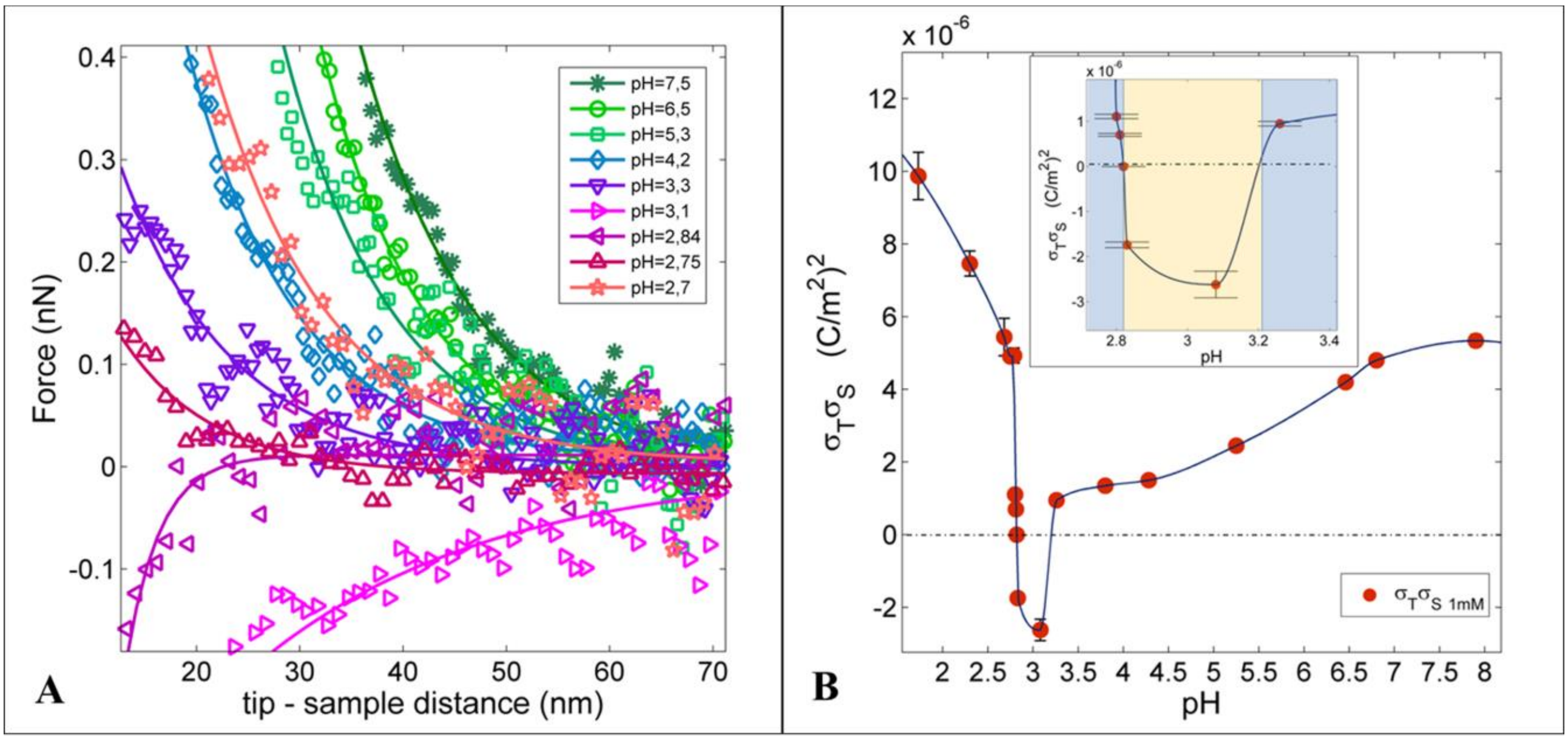

**Figure S7** (A) Force curves in 1mM NaCl at different pH values between the colloidal borosilicate glass tip and the borosilicate glass coverslip and (B) the $\sigma_S\sigma_T$ versus pH, extracted from the best fit of force curves. In the inset a magnification that clearly shows the two IEP of the surfaces.

The decreasing repulsion with pH corresponds to a decrease of the double layer interaction. At pH=3.08 the double-layer interaction becomes attractive and the value of Van der Waals interaction (independent of pH) is negligible compared to it (at a tip-surface distance of 14nm they are -0.1 nN and -0.02 nN respectively). The shift from repulsive to attractive double-layer interaction indicates that the IEP of colloidal glass tip lies between pH 3.3 and 3.1. When the pH of the solution is far enough from the IEP of the surface, the decay length is correctly described by the Debye Length of Eq. 4 (1mM NaCl at pH =3.81, $\lambda_D$= 9.49 nm).

In Fig. S7B is shown $\sigma_S\sigma_T$ extracted from the force curves as a function of pH and it is possible to identify immediately the IEP of the two surfaces. For high value of pH, both surfaces are negatively charged and so $\sigma_S\sigma_T$ is positive. Lowering the pH, we are approaching the first IEP of the system, and so the product of the surface charge density decreases until the zero value of the first surface IEP. When the pH value is lower than this first IEP value ($pH_{IEP}$ = 3.2), the charge density sign of one surface of the system changes and the interaction becomes attractive. The product of surface charge densities remains negative until the second IEP of the system is reached ($pH_{IEP}$=2.8). The slopes of the two positive regions of $\sigma_S\sigma_T$ versus pH are not equal. In fact, at lower pH, we are adding $10^{-2}$ M HCl, while at higher value of pH the amount of HCl is order of magnitude lower and the slope of the charging curve grows very slowly [13]. Despite the fact that both the colloidal probe and the glass coverslip used in this study are made of borosilicate glass, we found evidence of an asymmetric interface characterized by two different $pH_{IEP}$ values (Fig. S6B). The difference is small despite that fully resolved by our experimental apparatus ($pH_{IEP}$=3.20 ± 0.05 for the AFM probe vs $pH_{IEP}$= 2.82 ± 0.05 for the coverslip). The observed difference could be due to small changes in the relative abundances of silica and boron oxide components in borosilicate glasses, enhanced also by the different thermal annealing procedure and geometrical surface properties, which cause changes in the density of amphoteric sites (such compositional differences are in fact rather likely, due to batch-to-batch, as well as provider-to-provider fluctuations).

Assuming that the system is symmetric (a reasonable assumption due to the similarity of the IEPs of the probe and the substrate), and therefore $\sigma_T \approx \sqrt{\sigma_S \sigma_T}$ , we have calculated the net surface charge density of the AFM colloidal probe (Fig. S8).

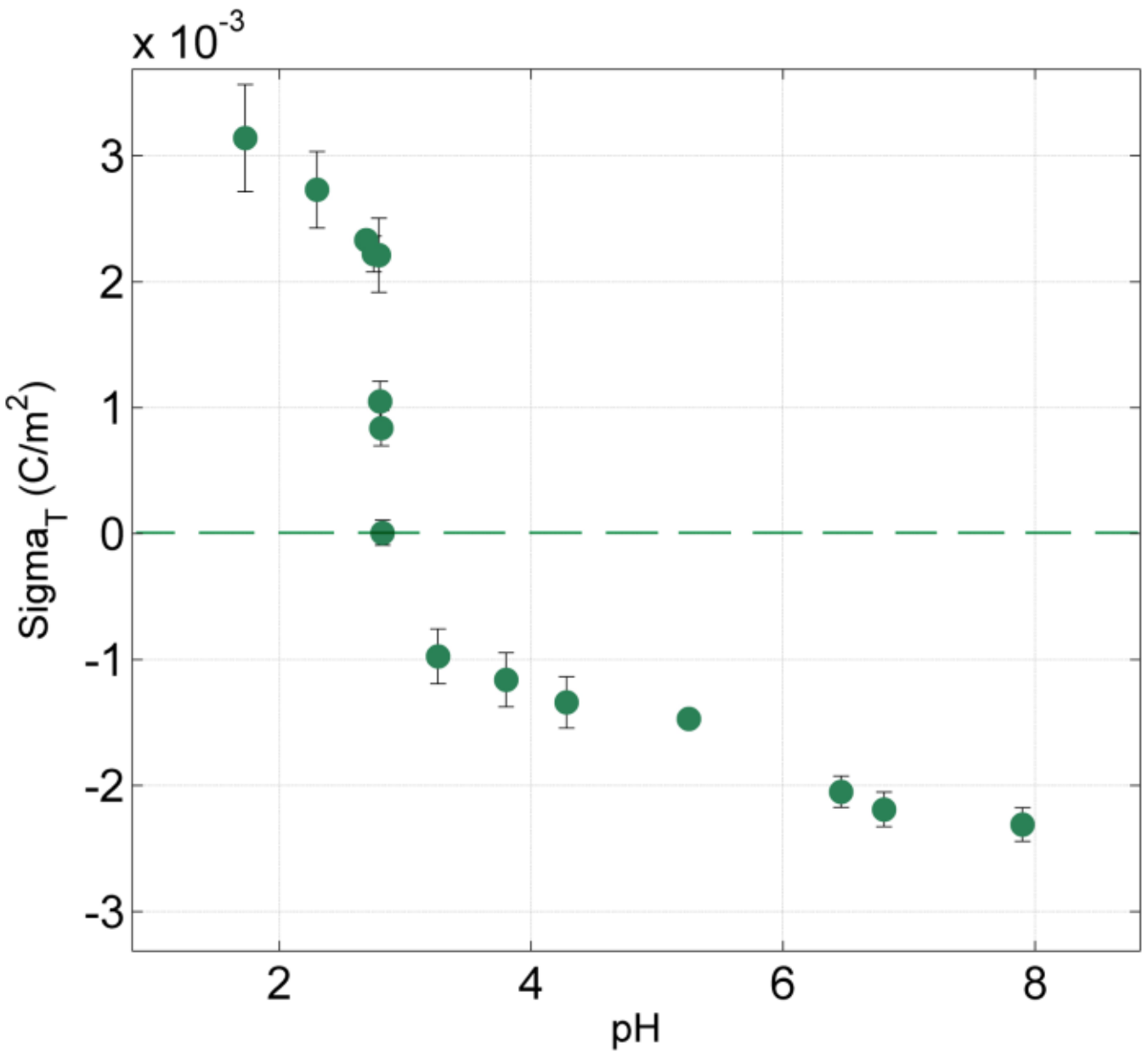

**Figure S8**. Net surface charge density of the AFM colloidal probe vs pH.

## 5.2.Single crystal <100> and polycrystalline rutile $TiO_2$

We have studied the interactions of the colloidal probe with reference single-crystal <100> and polycrystalline rutile $TiO_2$ surfaces (Table 2 in the main text). In the plots of $\sigma_S\sigma_T$ versus pH (Fig. S9B and S10B) it is possible to distinguish two different IEPs (one pertaining to the probe, the other to the sample). By comparing these plots, it is possible to determine precisely which one is the IEP of the probe; it recurred with high precision always in the same pH value for all the system studied, included the nanostructured ones (Fig. S11B-S19B). The attribution of $pH_{IEP}$ =3.2 value to the AFM probe was supported by the observation that this value is systematically measured in all experiments (which share the same borosilicate colloidal probe).

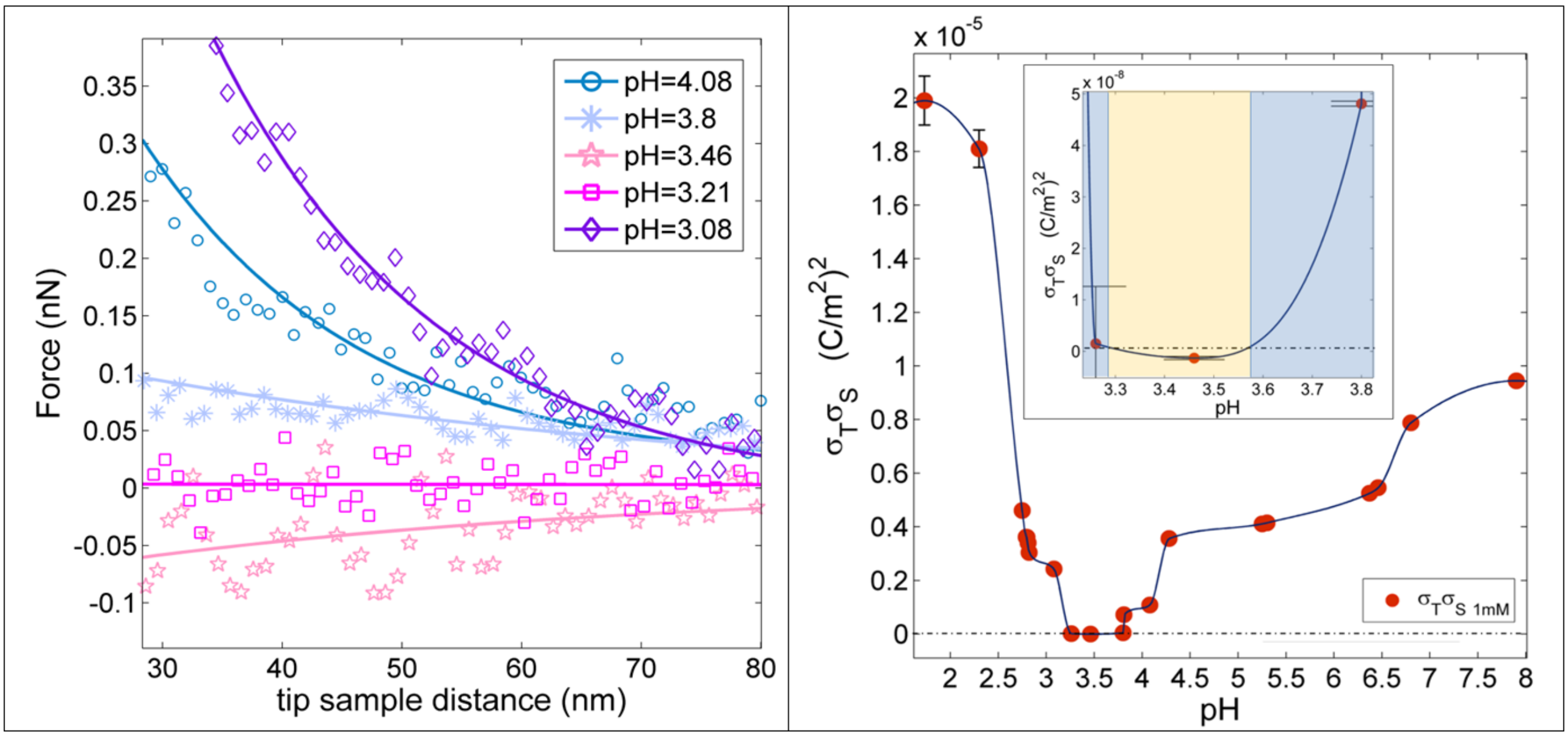

**Figure S9**. (Left) Force curves in 1mM NaCl at different pH values between the colloidal borosilicate glass tip and the Rutile flat $TiO_2$ substrate (crystallographic orientation <100>); (right) the $\sigma_S\sigma_T$ versus pH, extracted from the best fit of force curves. In the inset a magnification of the curve, which clearly identify the inversion of the charge sign due to the separation between the two IEPs of the system.

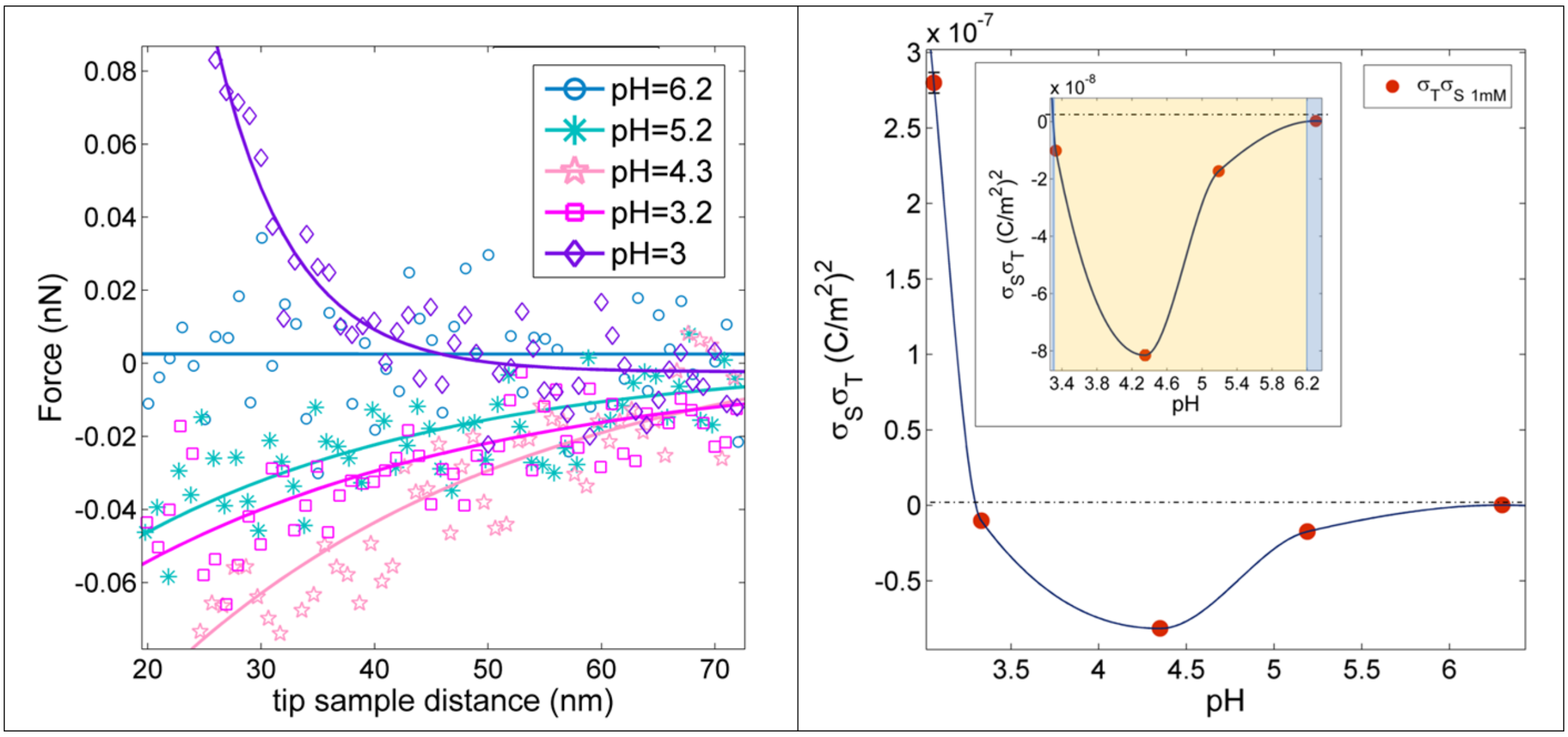

**Figure S10**. (Left) Force curves in 1mM NaCl at different pH values between the colloidal borosilicate glass tip and the Rutile flat polycrystalline $TiO_2$; (right) the $\sigma_S\sigma_T$ versus pH, extracted from the best fit of force curves. In the inset a magnification of the curve, which clearly identify the inversion of the charge sign due to the separation between the two IEPs of the system.

## 5.3. Nanostructured $TiO_2$

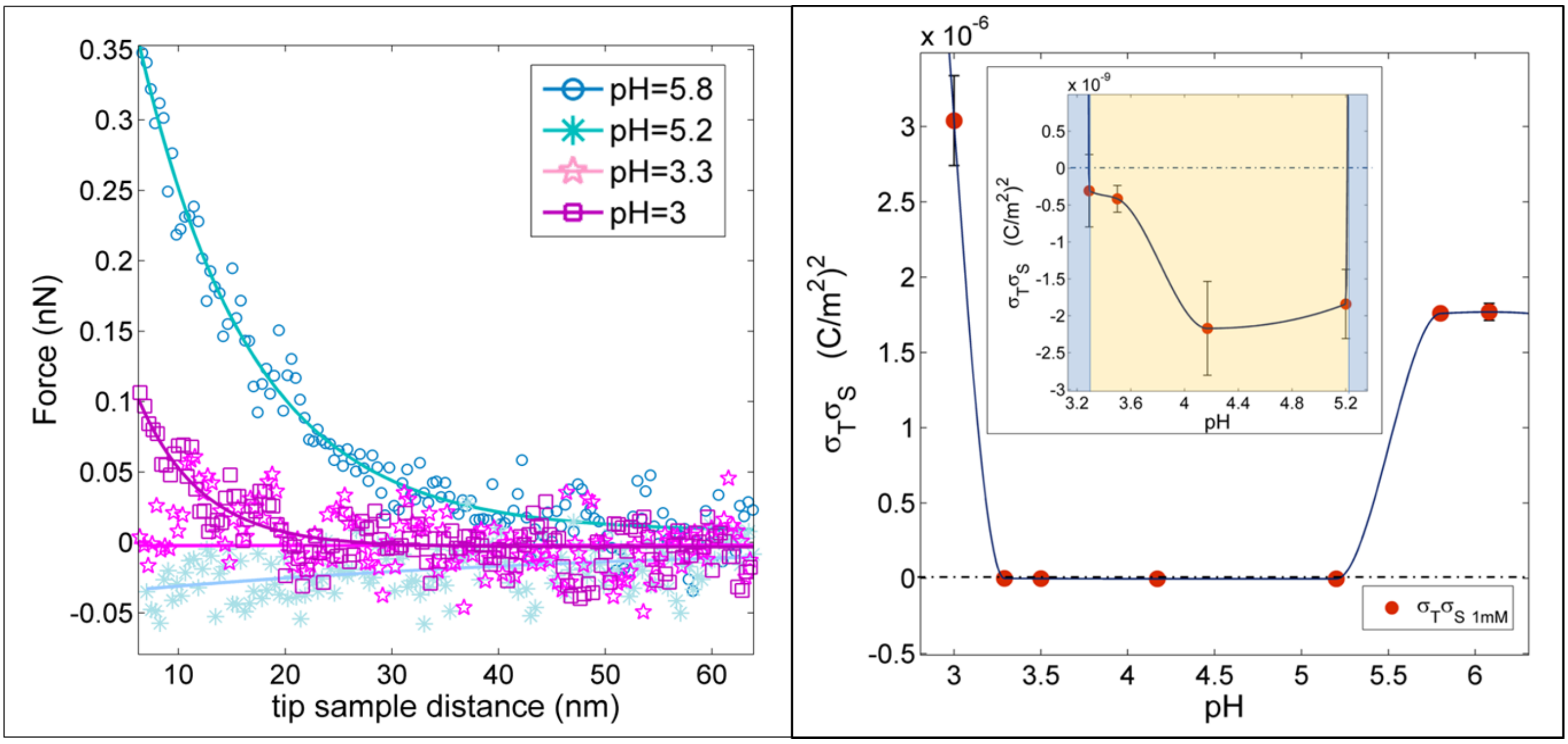

**Figure S11**. (Left) Force curves in 1mM NaCl at different pH values between the colloidal borosilicate glass tip and the rough ns-$TiO_2$ sample (Rq=5nm); (right) the $\sigma_S\sigma_T$ versus pH, extracted from the best fit of force curves. In the inset a magnification of the curve, which clearly identify the inversion of the charge sign due to the separation between the two IEPs of the system.

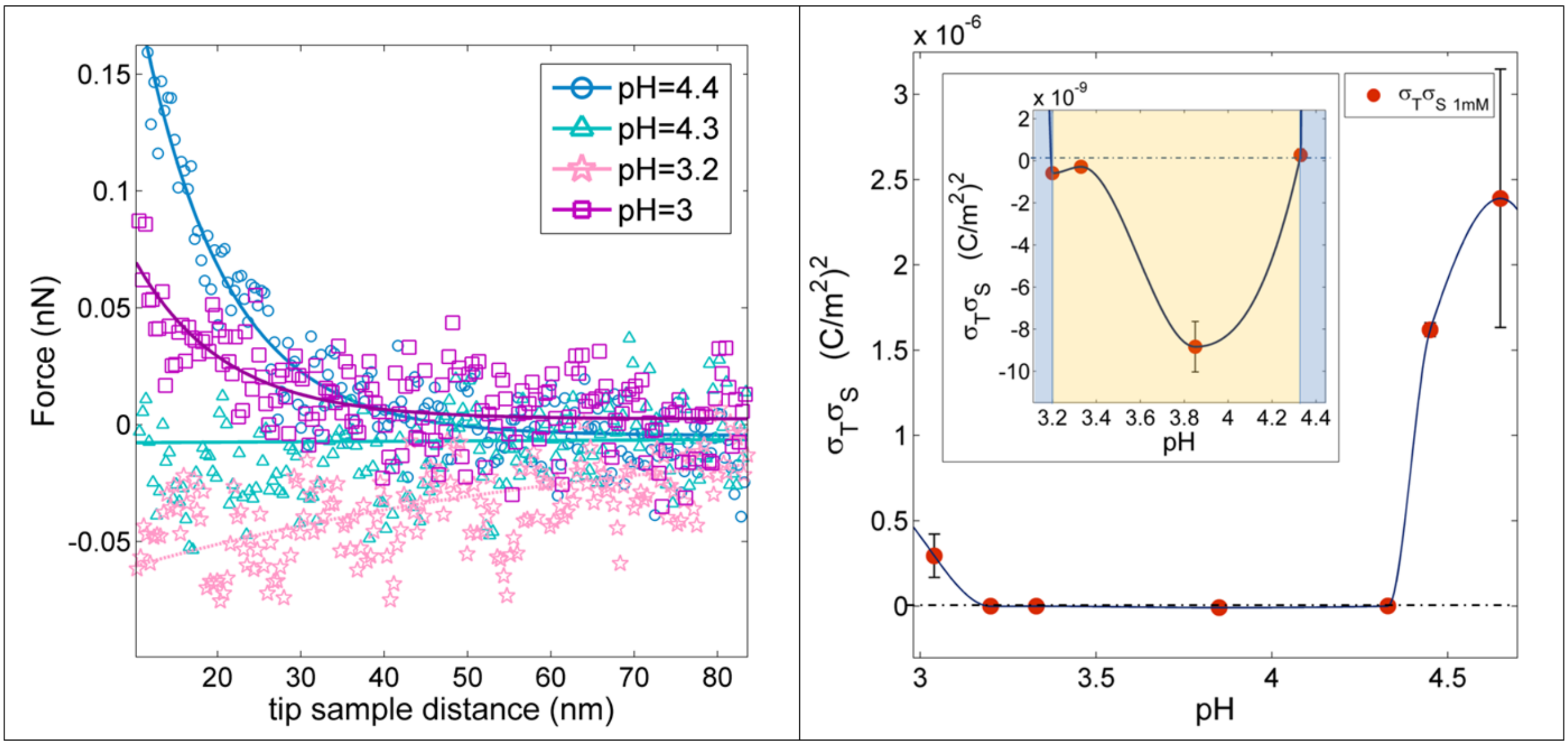

**Figure S12.** (Left) Force curves in 1mM NaCl at different pH values between the colloidal borosilicate glass tip and the rough ns-$TiO_2$ sample (Rq=10nm); (right) the $\sigma_S\sigma_T$ versus pH, extracted from the best fit of force curves. In the inset a magnification of the curve, which clearly identify the inversion of the charge sign due to the separation between the two IEPs of the system.

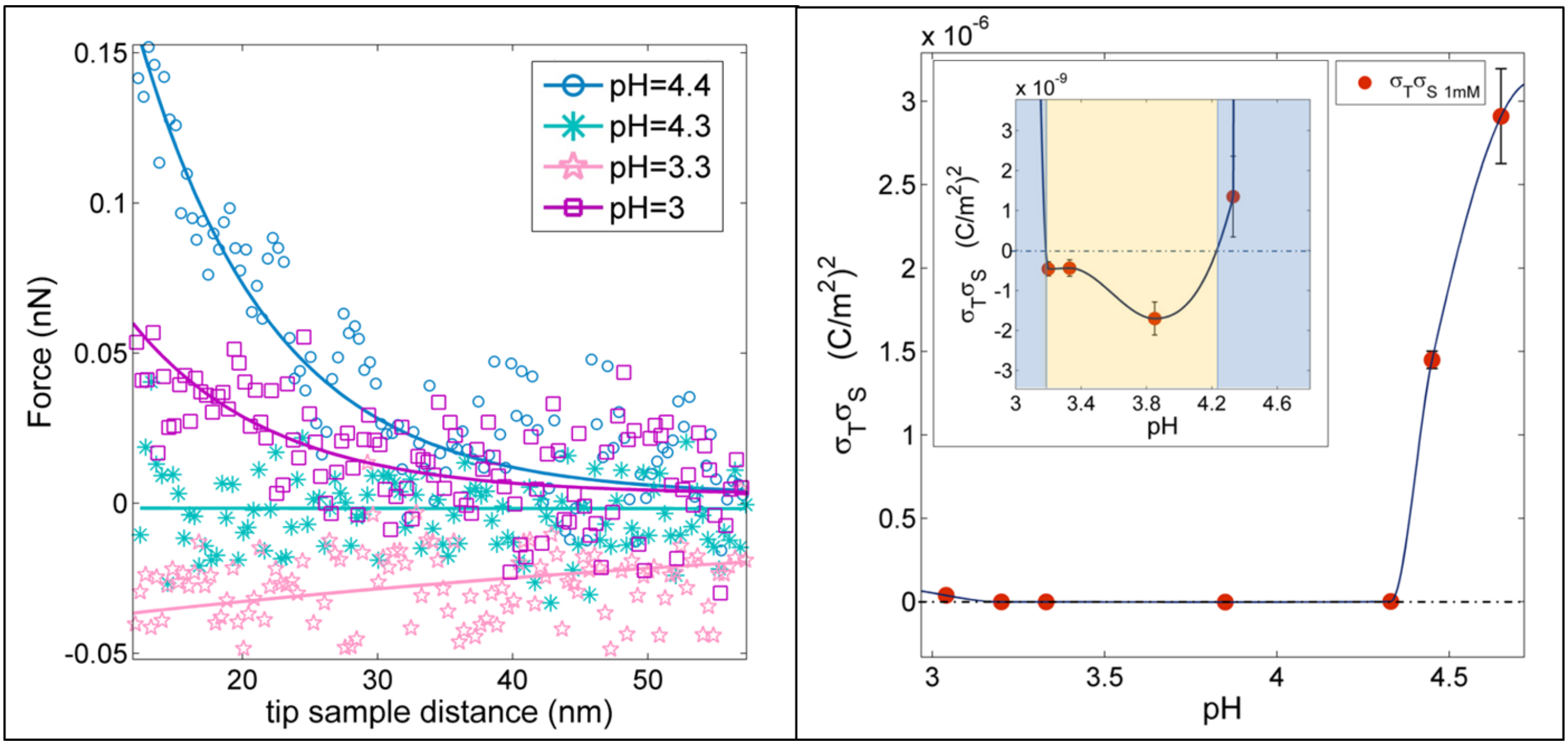

**Figure S13.** (Left) Force curves in 1mM NaCl at different pH values between the colloidal borosilicate glass tip and the rough ns-$TiO_2$ sample (Rq=14nm); (right) the $\sigma_S\sigma_T$ versus pH, extracted from the best fit of force curves. In the inset a magnification of the curve, which clearly identify the inversion of the charge sign due to the separation between the two IEPs of the system.

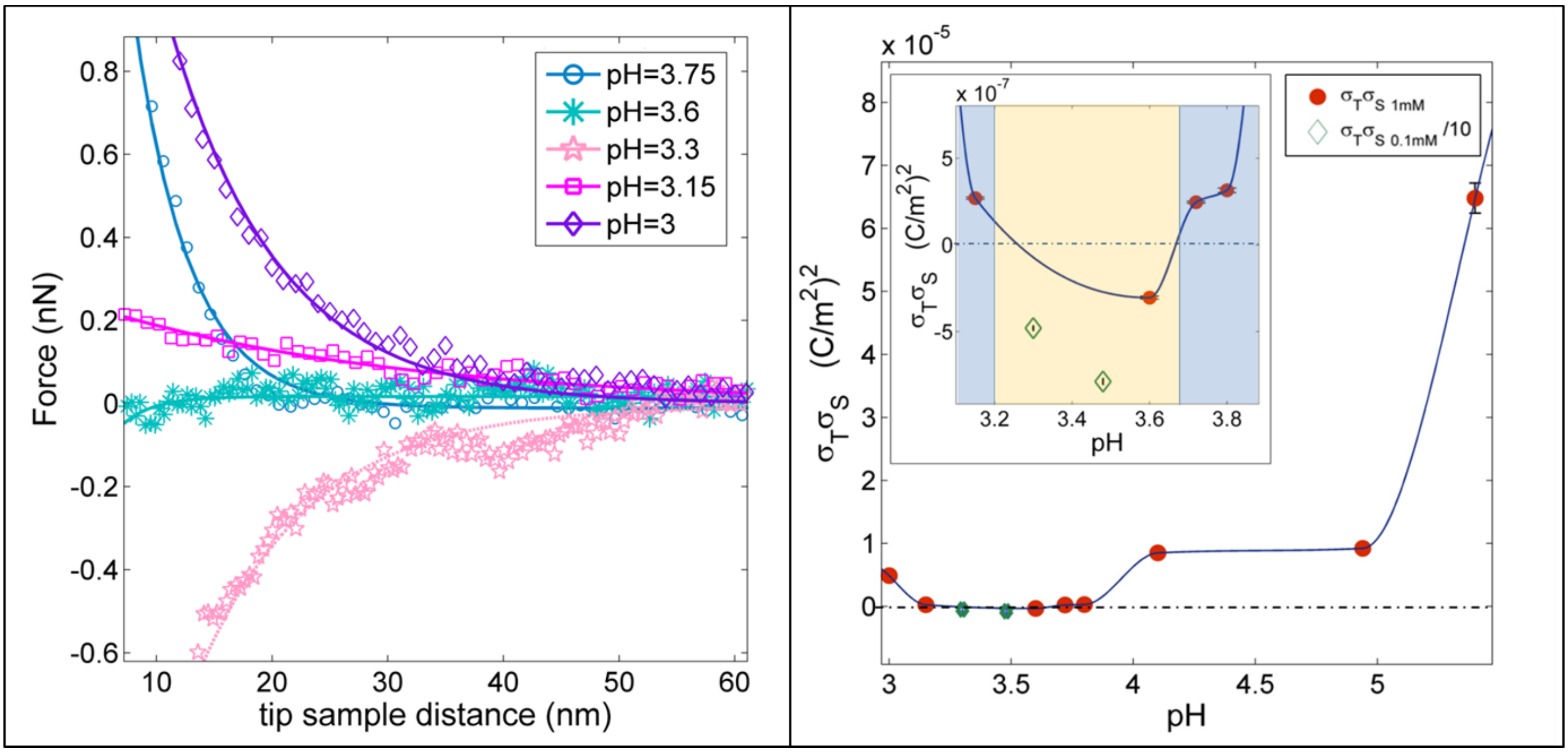

**Figure S14**. (Left) Force curves in 1mM NaCl at different pH values between the colloidal borosilicate glass tip and the rough ns-$TiO_2$ sample (Rq=17nm); (right) the $\sigma_S\sigma_T$ versus pH, extracted from the best fit of force curves. In the inset a magnification of the curve, which clearly identify the inversion of the charge sign due to the separation between the two IEPs of the system.

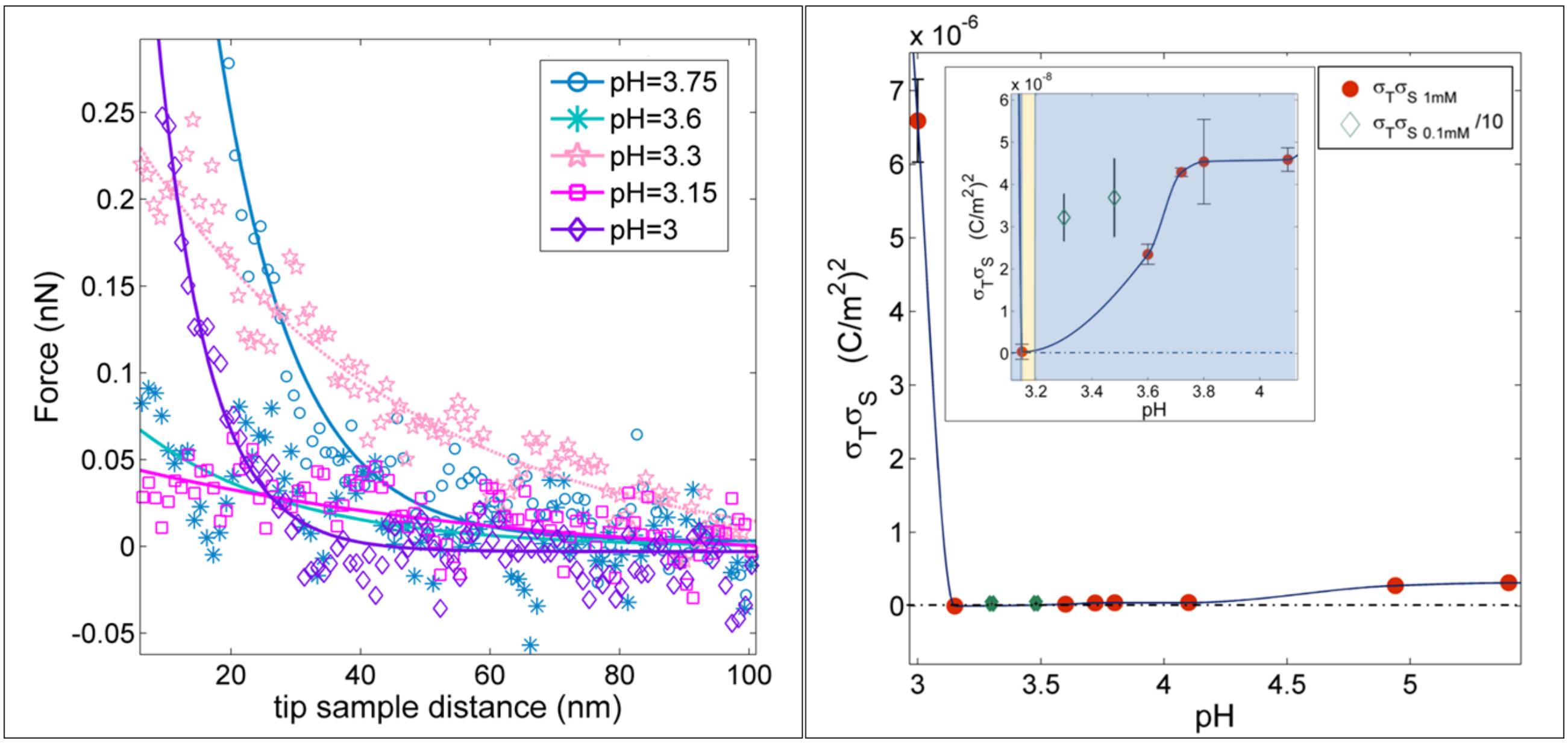

**Figure S15.** (Left) Force curves in 1mM NaCl at different pH values between the colloidal borosilicate glass tip and the rough ns-$TiO_2$ sample (Rq=19nm); (right) the $\sigma_S\sigma_T$ versus pH, extracted from the best fit of force curves. In the inset a magnification of the figure, which shows the overlapping between the two IEPs of the system.

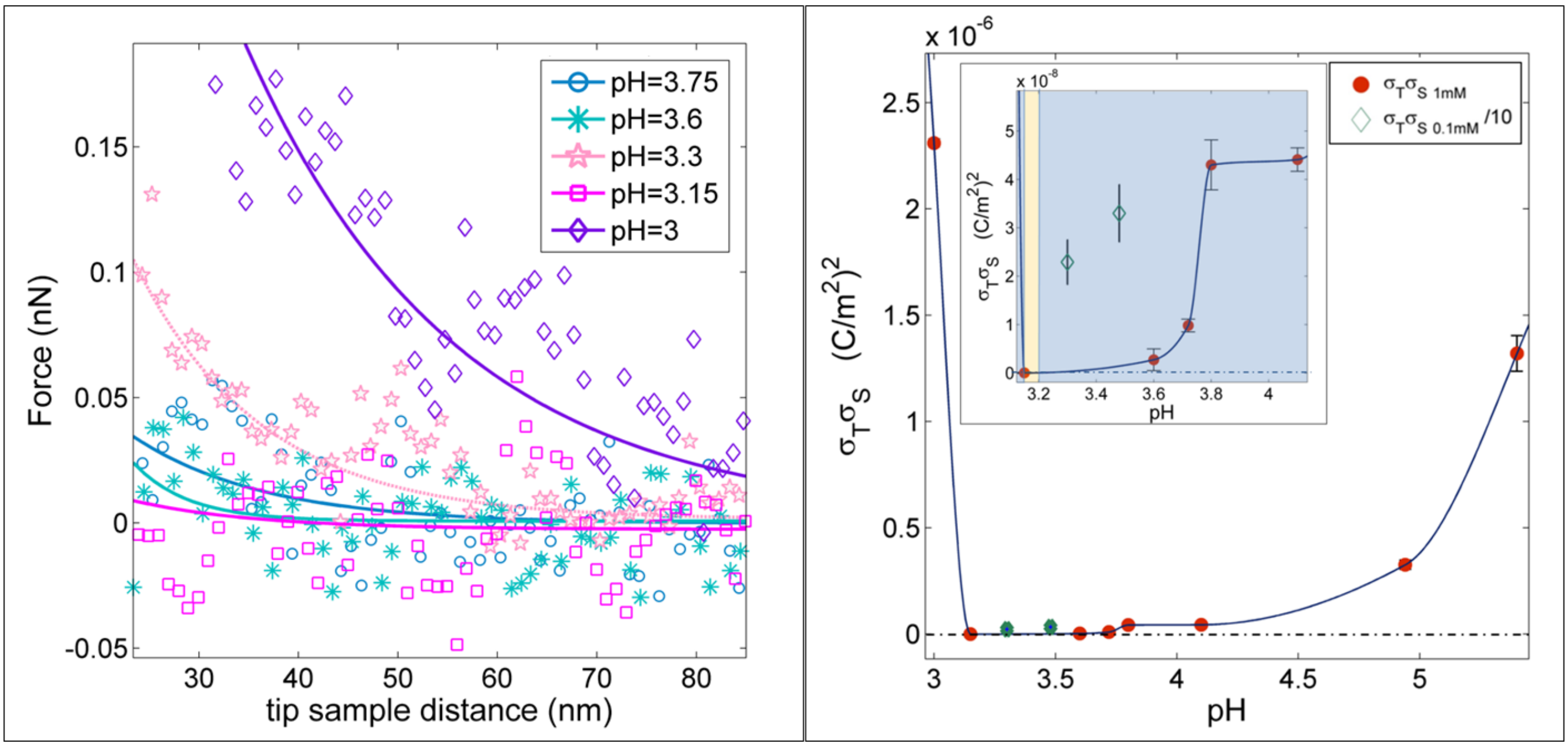

**Figure S16.** (Left) Force curves in 1mM NaCl at different pH values between the colloidal borosilicate glass tip and the rough ns-$TiO_2$ sample (Rq=20nm); (right) the $\sigma_S\sigma_T$ versus pH, extracted from the best fit of force curves. In the inset a magnification of the curve, which shows the overlapping between the two IEPs of the system.

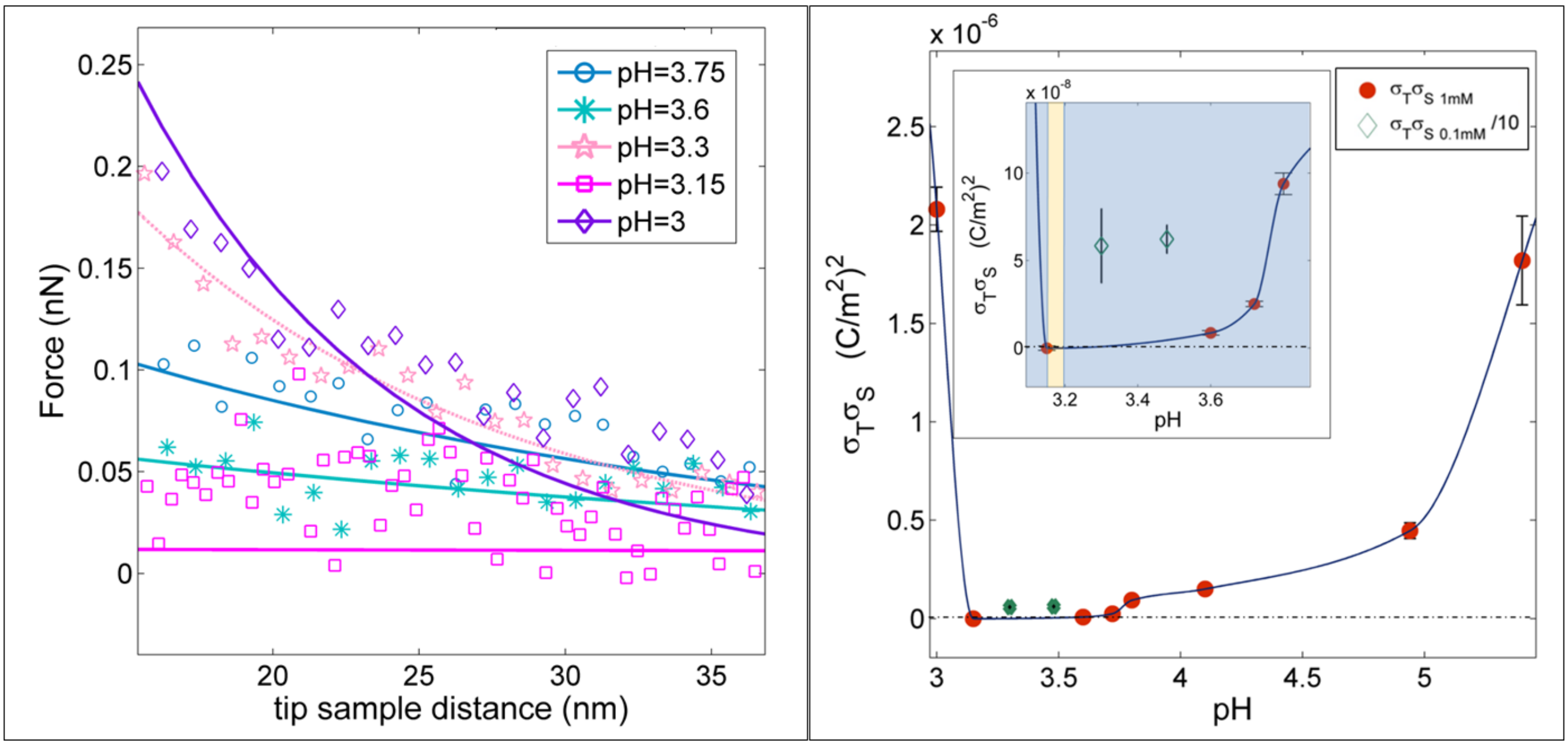

**Figure S17.** (Left) Force curves in 1mM NaCl at different pH values between the colloidal borosilicate glass tip and the rough ns-$TiO_2$ sample (Rq=21nm); (right) the $\sigma_S\sigma_T$ versus pH, extracted from the best fit of force curves. In the inset a magnification of the curve, which shows the overlapping between the two IEPs of the system.

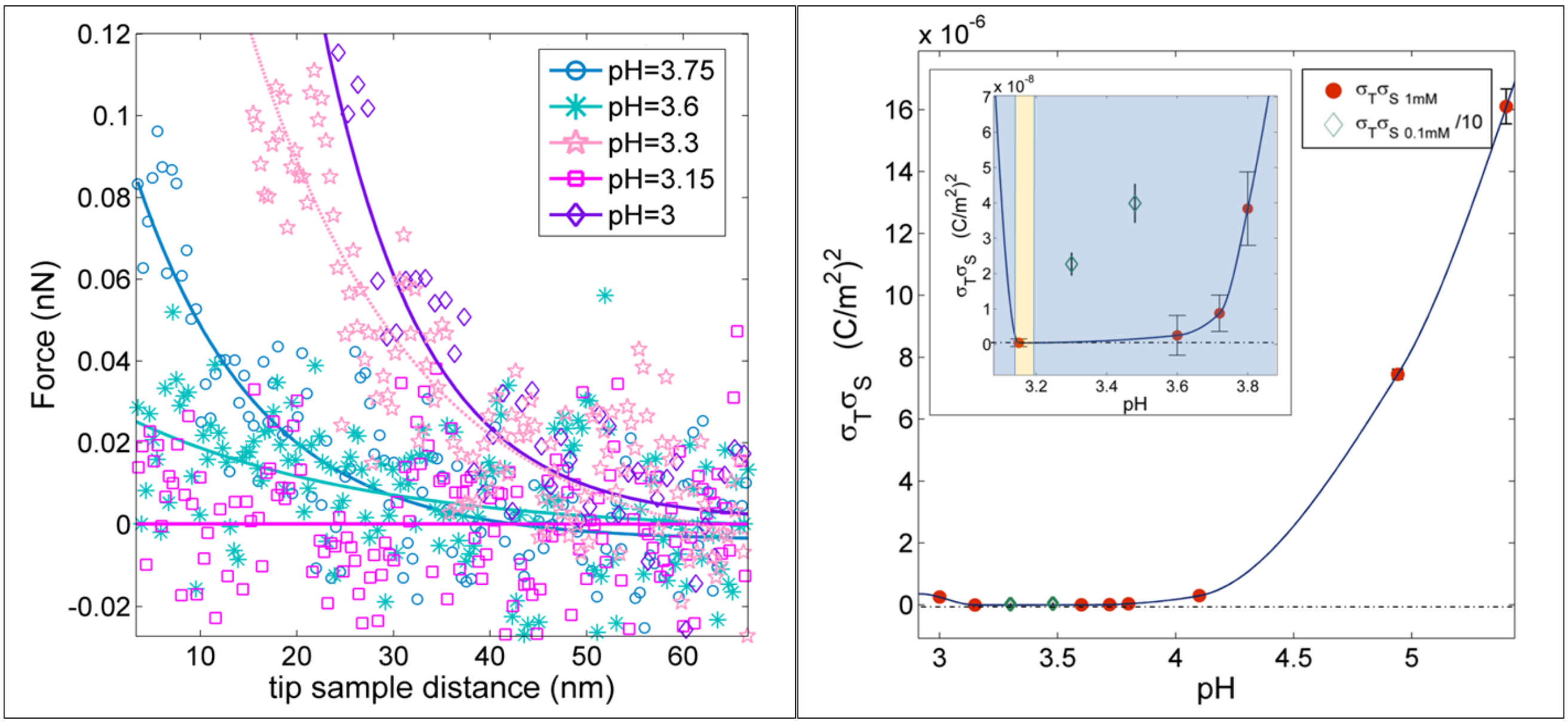

**Figure S18.** (Left) Force curves in 1mM NaCl at different pH values between the colloidal borosilicate glass tip and the rough ns-$TiO_2$ sample (Rq=22nm); (right) the $\sigma_S\sigma_T$ versus pH, extracted from the best fit of force curves. In the inset a magnification of the curve, which shows the overlapping between the two IEPs of the system.

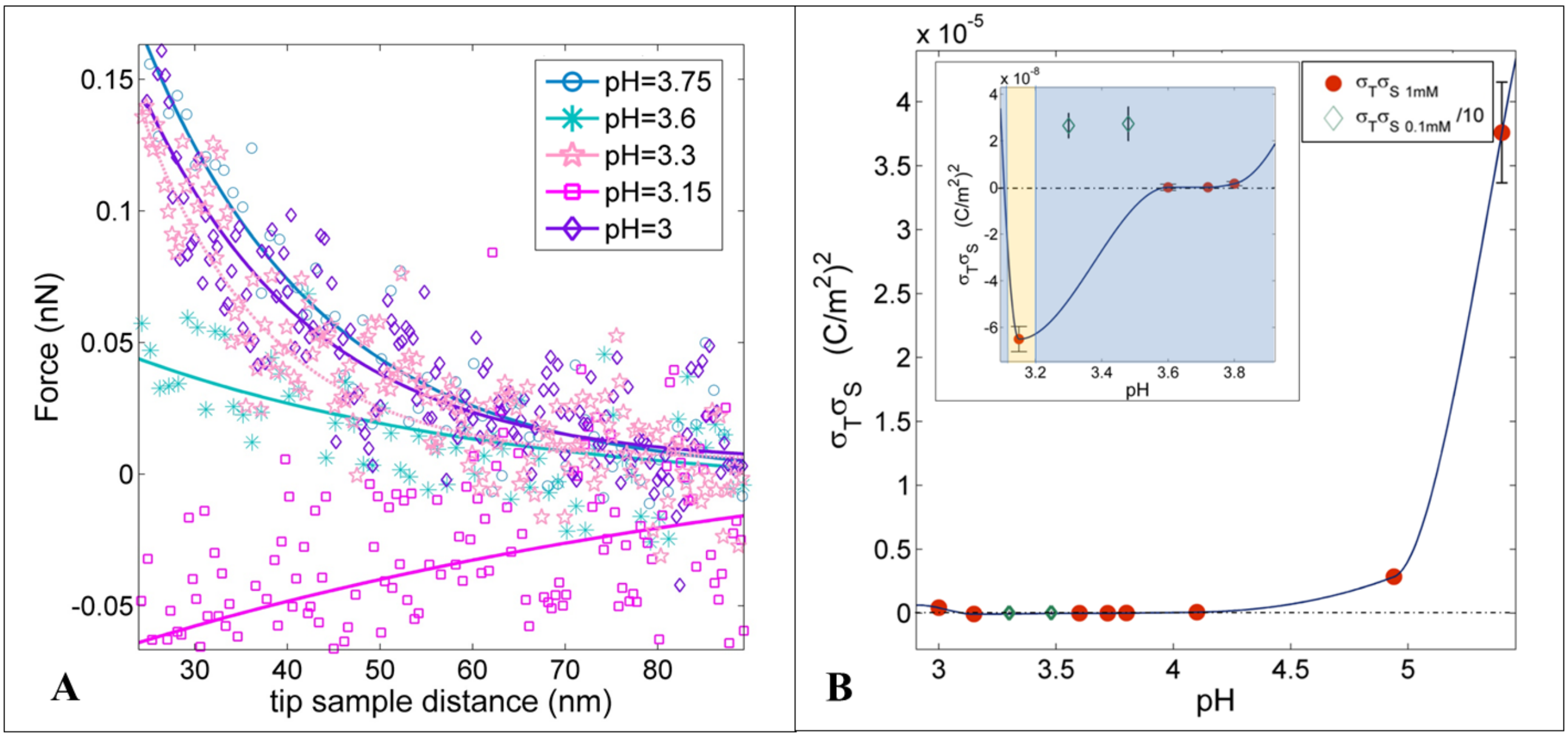

**Figure S19.** (Left) Force curves in 1mM NaCl at different pH values between the colloidal borosilicate glass tip and the rough ns-$TiO_2$ sample (Rq=26nm); (right) the $\sigma_S\sigma_T$ versus pH, extracted from the best fit of force curves. In the inset a magnification of the curve, which shows the overlap between the two IEPs of the system.

## 6. Self-overlap of electrostatic double-layers: a simplified picture

The double layer structure is assumed to consist in a volume of depth $\lambda_D$ stemming perpendicularly from the solid surface toward the bulk of the electrolyte.

We consider as the total double layer volume of a single pore the sum of the two regions originating from the two slopes of the pore. We consider a 2-dimensional projection of the pore, so that the volume of the double layer is in fact an area $\Sigma_0$. Our results should be the same, apart from a multiplicative factor, in the 3-dim. case.

The pore has slope $\tan(\theta)=2R_q/\xi$ (see Fig. S20).

The area of the overlapping region is $\Sigma$. We introduce the self-overlap parameter $\gamma=\Sigma/\Sigma_0$.

We distinguish between two cases: $\theta\leq 45°$ ($2R_q/\xi\leq 1$), and $\theta>45°$ ($2R_q/\xi>1$). We will calculate the ratio $\gamma$ only for $\lambda_D<\lambda^*_D$. $\lambda^*_D$ represents the depth of the double layer at which the shape of the overlapping regin changes from a quadrilateral (a kite, for $\theta\leq 45°$, or a rhombus, for $\theta>45°$), from a more complex polygon. $\lambda^*_D$ is shown for the two cases in Fig. S20-left and S21-left.

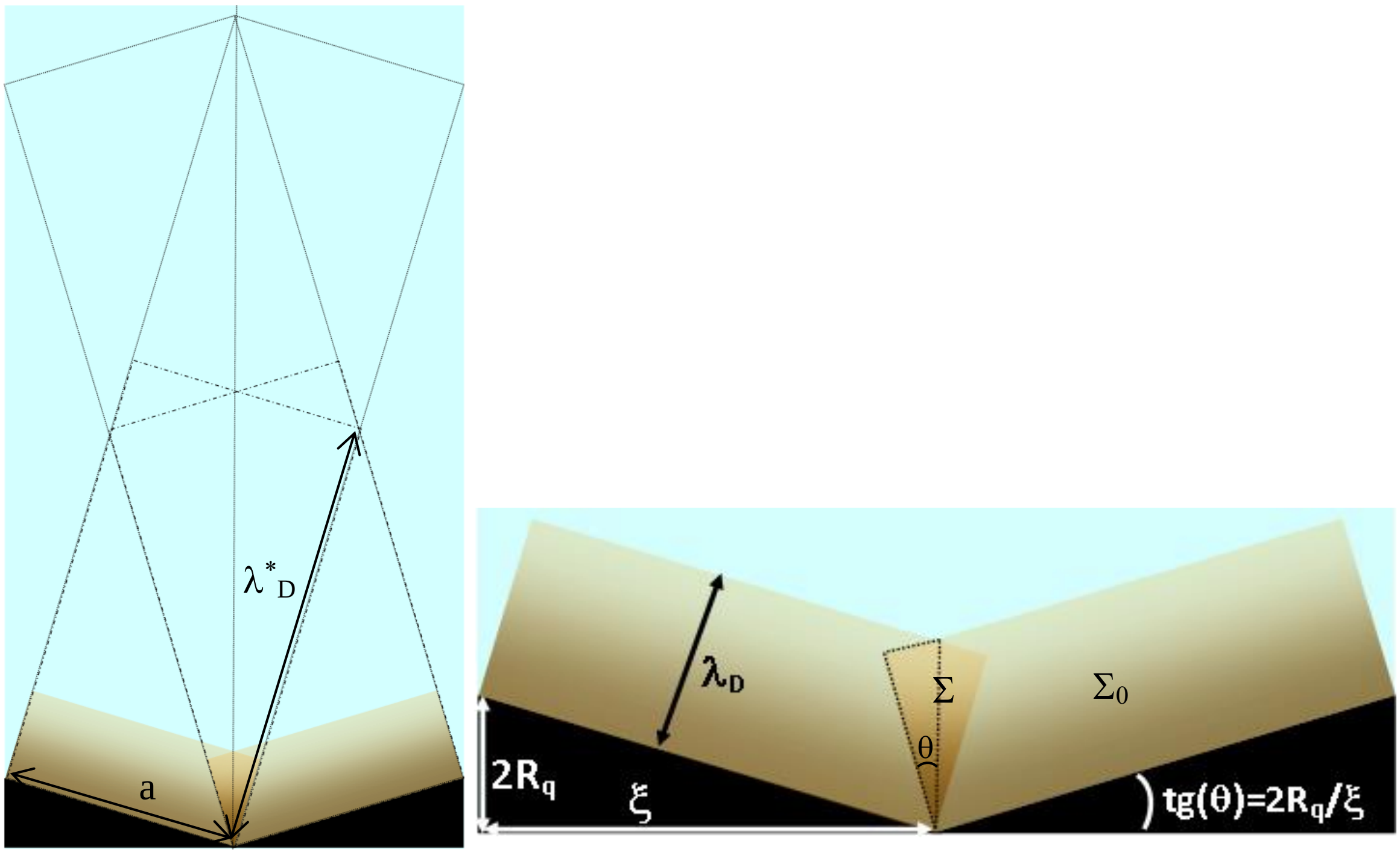

**Figure S20.** The simplified double layer structure of a surface pore, in the case $\theta\leq 45°$ ($2R_q/\xi\leq 1$). On the left, it can be seen that the shape of the overlapping region is a kite of area $\Sigma$ for $\lambda_D<\lambda^*_D$. On the right, a magnified view of the double layer structure for $\lambda_D<\lambda^*_D$.

***Case I, $\theta \leq 45°$ ($2R_q/\xi \leq 1$) and $\lambda_D \leq \lambda^*_D$***

From Fig. S19-left it follows that:

$$\lambda_D^* \tan\theta = \frac{1}{2}\xi\sqrt{1+\left(\frac{2R_q}{\xi}\right)^2}$$

being $a = \xi\sqrt{1+\left(\frac{2R_q}{\xi}\right)^2}$,

so that:

$$\lambda_D^* = \frac{\xi\sqrt{1+\left(\frac{2R_q}{\xi}\right)^2}}{2\left(\frac{2R_q}{\xi}\right)}, \tag{S5}$$

In the case of $2R_q/\xi << 1$, the condition $\lambda_D \leq \lambda^*_D$ holds for $\lambda_D$ up to several times larger than $\xi$.

$\Sigma$ is twice the area of the right triangle highlighted by the dotted line in Fig. S20-right:

$$\Sigma = \lambda_D^2\left(\frac{2R_q}{\xi}\right), \tag{S6}$$

$\Sigma_0$ is twice the area of the double layer of each pore wall $a\lambda_D$ minus the area $\Sigma$ common overlap region. It follows that:

$$\Sigma_0 = 2\lambda_D\xi\sqrt{1+\left(\frac{2R_q}{\xi}\right)^2} - \lambda_D^2\left(\frac{2R_q}{\xi}\right), \tag{S7}$$

Eventually:

$$\gamma = \frac{\left(\frac{\lambda_D}{\xi}\right)\left(\frac{2R_q}{\xi}\right)}{2\sqrt{1+\left(\frac{2R_q}{\xi}\right)^2} - \left(\frac{\lambda_D}{\xi}\right)\left(\frac{2R_q}{\xi}\right)}, \tag{S8}$$

At $\lambda_D = \lambda^*_D$, $\gamma = 1/3$.

***Case II, θ>45° (2R$_q$/ξ>1) and $\lambda_D \leq \lambda^*_D$***

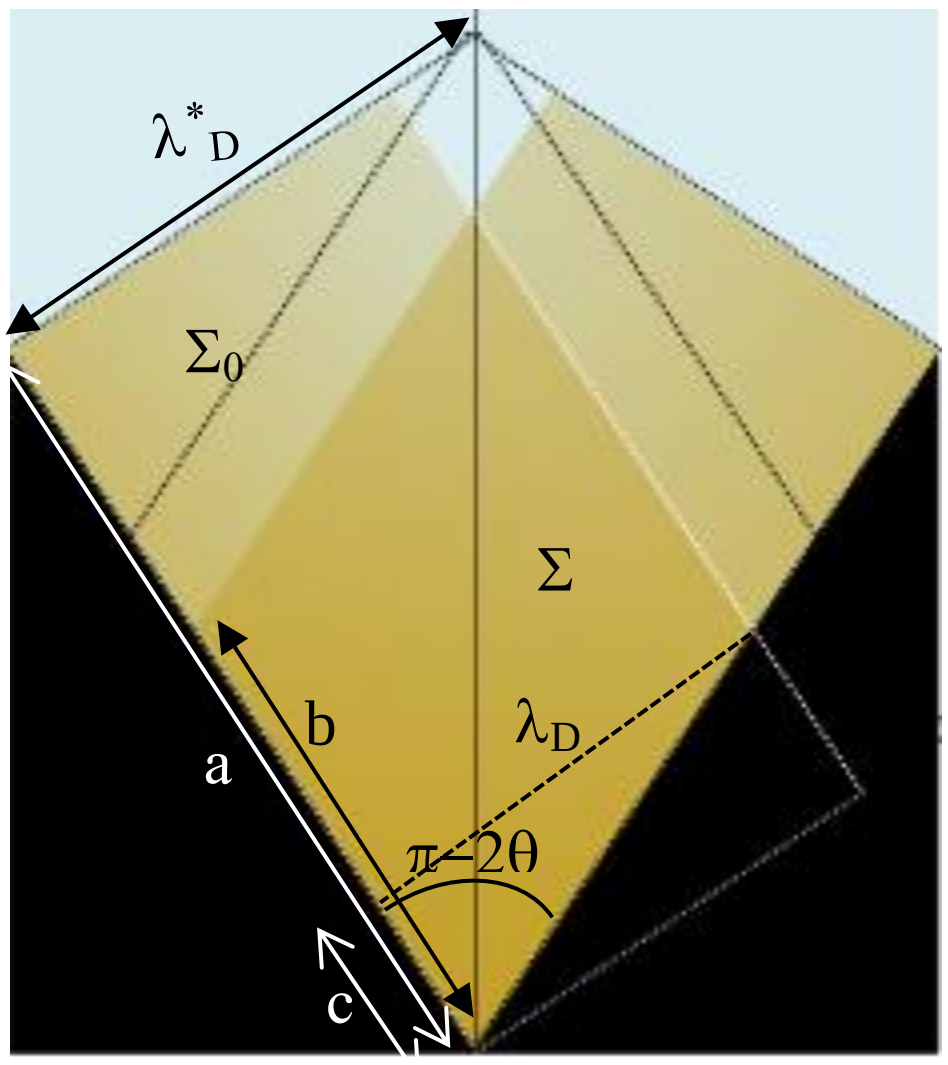

**Figure S21.** The simplified double layer structure of a surface pore, in the case θ>45° (2R$_q$/ξ>1). The overlapping region is a rhombus for $\lambda_D<\lambda^*_D$.

$$\lambda_D^* = \frac{\xi\sqrt{1+\left(\frac{2R_q}{\xi}\right)^2}}{\left(\frac{2R_q}{\xi}\right)}$$

In the case of 2R$_q$/ξ>1, the condition $\lambda_D \leq \lambda^*_D$ holds for $\lambda_D$ up to about 1.5ξ.

$\Sigma = b\lambda_D,$

$b = \frac{\lambda_D}{\sin(\pi-2\theta)} = \lambda_D/\sin(2\theta) = \lambda_D/2\sin(\theta)\cos(\theta) = \frac{\lambda_D}{2\tan\theta}[1+(\tan\theta)^2],$

therefore:

$$\Sigma = \frac{\lambda_D^2}{2}\frac{1+\left(\frac{2R_q}{\xi}\right)^2}{\left(\frac{2R_q}{\xi}\right)}.$$

$\Sigma_0 = 2a\lambda_D - 2\left(\frac{\lambda_D c}{2}\right)$- Σ,

where $c = \frac{\lambda_D}{\tan(\pi-2\theta)}$ and $\tan(\pi-2\theta) = \frac{2\tan\theta}{(\tan\theta)^2-1}$.

It follows:

$$\Sigma_0 = 2\lambda_D \xi \sqrt{1 + \left(\frac{2R_q}{\xi}\right)^2 - \lambda_D^2 \left(\frac{2R_q}{\xi}\right)}$$

as in the case θ≤45°.

$$\gamma = \frac{\left(\frac{\lambda_D}{\xi}\right)\left(\frac{2R_q}{\xi}\right)}{2\sqrt{1+\left(\frac{2R_q}{\xi}\right)^2 - \left(\frac{\lambda_D}{\xi}\right)\left(\frac{2R_q}{\xi}\right)}} \frac{1}{2}\left[1 + \frac{1}{\left(\frac{2R_q}{\xi}\right)^2}\right], \tag{S9}$$

At $\lambda_D = \lambda^*_D$, $\gamma >> 0.5$.